\pdfoutput=1
\documentclass[12pt,a4paper]{article}
\usepackage{jheplike,mathtools,todonotes,tikz}

\hypersetup{pdftitle=Banana,pdfcreator={},
            linkcolor=[rgb]{0.15,0.35,0.75},
            colorlinks=true,citecolor=[rgb]{0.675,0,0.2},
            urlcolor=[rgb]{0.15,0.35,0.65}}

\newcommand{\eqs}[1]{\begin{gather}\begin{aligned}#1\end{aligned}\end{gather}}
\newcommand{\ddl}{\mathfrak{D}^d\ell}

\newcommand{\cI}{\mathcal{I}}
\newcommand{\cL}{\mathcal{L}}
\newcommand{\cW}{\mathcal{W}}
\newcommand{\cM}{\mathcal{M}}
\newcommand{\cO}{\mathcal{O}}
\newcommand{\cS}{\mathcal{S}}
\newcommand{\cT}{\mathcal{T}}

\newcommand{\eps}{\epsilon}
\newcommand{\ivec}{\begin{pmatrix}\cI_1(\eps;x)\\\cI_2(\eps;x)\\\cI_3(\eps;x)\end{pmatrix}}

\newcommand{\sm}[2]{\begin{smallmatrix}4 & 6 \\ #1 & #2\end{smallmatrix}}
\newcommand{\smz}{\begin{smallmatrix}0 & 0 \\ 0 & 0\end{smallmatrix}}

\newcommand{\sGt}{\tilde{\Gamma}}

\newcommand{\sGtAS}{\Gamma_{\text{AS}}}
\newcommand{\smg}[1]{\begin{smallmatrix} #1\end{smallmatrix}}

\makeatletter
\g@addto@macro\bfseries{\boldmath}
\makeatother

\newif\ifnote 
\notetrue

\allowdisplaybreaks

\def\beq{\begin{equation}}
\def\eeq{\end{equation}}
\def\bsp#1\esp{\begin{split}#1\end{split}}

\DeclareMathOperator{\K}{K}
\DeclareMathOperator{\E}{E}
\DeclareMathOperator{\SL}{SL}
\newcommand{\mymod}{{\hskip -0.2cm}\mod}
\newcommand{\IMF}[2]{\mathcal{I}\hskip-0.08cm\left(\begin{smallmatrix} #1\end{smallmatrix};#2\right)}

\title{An analytic solution for the equal-mass banana graph}

\author[a]{Johannes Broedel,} 
\author[b,c]{Claude Duhr,}
\author[d]{Falko Dulat,}
\author[c]{Robin Marzucca,}
\author[b]{Brenda Penante,}
\author[b]{Lorenzo Tancredi}

\affiliation[a]{Institut f\"{u}r Mathematik und Institut f\"{u}r Physik,
Humboldt-Universit\"{a}t zu Berlin,\\
IRIS Adlershof, Zum Grossen Windkanal 6, 12489 Berlin, Germany} 
\affiliation[b]{Theoretical Physics Department, CERN, Geneva, Switzerland} 
\affiliation[c]{Center for Cosmology, Particle Physics and Phenomenology (CP3),\\
Universit\'e Catholique de Louvain, 1348 Louvain-La-Neuve, Belgium}
\affiliation[d]{SLAC National Accelerator Laboratory, Stanford University, Stanford, CA 94309, USA}

\emailAdd{jbroedel@physik.hu-berlin.de}
\emailAdd{claude.duhr@cern.ch}
\emailAdd{dulatf@slac.stanford.edu}
\emailAdd{robin.marzucca@uclouvain.be}
\emailAdd{b.penante@cern.ch}
\emailAdd{lorenzo.tancredi@cern.ch}

\abstract{We present fully analytic results for all master integrals for the
three-loop banana graph with four equal and non-zero masses.  The results are
remarkably simple and all integrals are expressed as linear combinations of
iterated integrals of modular forms of uniform weight for the same congruence
subgroup as for the two-loop equal-mass sunrise graph. We also show how to
write the results in terms of elliptic polylogarithms evaluated at rational
points.}

\preprint{\begin{minipage}[t]{8cm}\begin{flushright}CP3-19-34, CERN-TH-2019-105\\
            HU-Mathematik-2019-04, HU-EP-19/20\\
            SLAC-PUB-17453\end{flushright}\end{minipage}}

\begin{document}
\maketitle\thispagestyle{empty}



\section{Introduction}

Feynman integrals are the most important building blocks required to study
scattering processes in perturbative quantum field theory. 
The physics program at the Large Hadron Collider has benefitted dramatically
from the availability of theoretical predictions with high degree of accuracy.
These predictions were made possible by tremendous advancements in the
calculation of multi-loop scattering amplitudes in recent
years.
However, it has become clear that future efforts to further test our theoretical
understanding of the nature of particle interactions at high energies will
require an even higher level of precision. Computing scattering processes to
even higher orders in perturbation theory will therefore require a deeper understanding of
multi-loop Feynman integrals and a further refinement of the mathematical
technology used to evaluate them.

Feynman integrals encode the complicated branch cut structure of scattering
amplitudes, reflecting the structure of physical thresholds of scattering
processes. 
Consequently, Feynman integrals need to be described in terms of classes of
special functions that exhibit the required branch cuts. The classic examples are
the logarithm and dilogarithm functions that encode the branch cut structure of simple
one-loop amplitudes in four space-time dimensions. More complicated Feynman integrals require
functions with a richer analytical structure in order to properly encode their branch cut structure.
In this context,
multiple polylogarithms~\cite{Kummer, Nielsen, Goncharov:1998kja} have proven 
an amazingly successful class of
functions to describe many scattering processes, in particular in phenomenologically 
interesting cases where  no massive particles circulate inside the loops.

However, it is well known that MPLs do not exhaust the space of functions to which Feynman integrals evaluate. It particular, it has been known for several decades that starting from two loops not all Feynman integrals evaluate to MPLs~\cite{Broadhurst:1987ei,Bauberger:1994by,Bauberger:1994hx,Laporta:2004rb,Kniehl:2005bc,Aglietti:2007as,Czakon:2008ii,Brown:2010bw,MullerStach:2011ru,CaronHuot:2012ab,Huang:2013kh,Brown:2013hda,Nandan:2013ip}, though no complete analytic results were known. 
This situation has changed with the work of Bloch and Vanhove~\cite{Bloch:2013tra}, who have shown that the simplest example of a Feynman integral that cannot be evaluated in terms of MPLs is in fact expressible through a generalisation of the dilogarithm to an elliptic curve. This result has sparked a lot of activity over the last few years, and by now we have complete analytic results for many Feynman integrals that involve functions of elliptic type~\cite{Adams:2013nia,Adams:2014vja,Adams:2015gva,Adams:2015ydq,Remiddi:2016gno,Primo:2016ebd,Bonciani:2016qxi,Adams:2016xah,Passarino:2016zcd,vonManteuffel:2017hms,Ablinger:2017bjx,Chen:2017pyi,Hidding:2017jkk,Bogner:2017vim,Bourjaily:2017bsb,Broedel:2017siw,Laporta:2017okg,Broedel:2018iwv,Broedel:2018qkq,Adams:2018bsn,Adams:2018kez,Broedel:2019hyg,Bogner:2019lfa}. In all cases these results involve new classes of transcendental functions, related either to elliptic generalisations of MPLs~\cite{Bloch:2013tra,BrownLevin,Remiddi:2017har,Broedel:2017kkb} or iterated integrals of modular forms~\cite{ManinModular,Brown:mmv,Adams:2017ejb,Broedel:2018iwv}. Incidentally, these are also the same class of functions which describe string amplitudes at genus one~\cite{Broedel:2014vla,Broedel:2015hia,Broedel:2017jdo,Broedel:2018izr}.

It is also known that functions related to more complicated geometries show
up~\cite{Brown:2010bw,Huang:2013kh,Bloch:2014qca,Bloch:2016izu,Broadhurst:2016myo,Bourjaily:2018yfy,Bourjaily:2018ycu}.
The simplest example of such an integral is probably the three-loop banana graph
with four massive propagators, whose associated geometry is a specific family of
K3 surfaces~\cite{Bloch:2014qca}. While functions of elliptic type that arise in
Feynman integral computations start to be well understood, we still lack
a clear picture of the class of functions that arise from more complicated
geometries. Hence, no complete analytic results are known for the banana graph
in terms of a well-defined class of transcendental functions. 

An exception to this case is the limit where all four masses in the banana
graph are equal. In this case the K3 surface is elliptically fibered, and the
base and the fiber are described by the same elliptic curve. This elliptic
curve, in turn, is related to the elliptic curve of the sunrise
integral~\cite{Bloch:2014qca}. The corresponding family of K3 surfaces and
their associated Picard-Fuchs operator were studied in ref.~\cite{verrill1996},
where it was shown that the solutions of this operator can be written in terms
of the solutions of the Picard-Fuchs operator of the sunrise graph. In
ref.~\cite{Primo:2017ipr} this property was used to express all master
integrals for the equal-mass banana graph in terms of iterated integrals whose
integration kernels involve products of complete elliptic integrals.  However,
a complete analytic solution of all master integrals for the equal-mass banana
graph in terms of a well-defined and well-studied class of functions is
currently still lacking.

In the remainder of this paper we close this gap and we present for the first
time complete analytic results for all three master integrals of the equal-mass
three-loop banana graph in $d=2$ dimensions.  Our starting point is the
differential equation of refs.~\cite{verrill1996,Primo:2017ipr}.  From there we
show that, since the homogeneous solutions can be expressed in terms of the
same modular forms that appear in the computation of the sunrise graph, the
differential equation for the master integrals of the banana graph can be
solved in terms of the same class of functions as for the sunrise graph. When
expressed in this way, our results are characterised by a remarkable
simplicity.  Moreover, we observe that all master integrals can be written as
linear combinations of pure functions of uniform weight, as defined in
ref.~\cite{Broedel:2018qkq}.

The paper is organised as follows: in section~\ref{sec:bananareview} we review
the banana graph, its differential equations and the results of
refs.~\cite{Bloch:2014qca} and \cite{Primo:2017ipr}.  In
section~\ref{sec:sunrise_review} we illustrate how to solve the differential
equation of the sunrise graph in terms of iterated integrals of modular forms,
and we introduce the relevant class of functions.  In
section~\ref{sec:bananaforms} we present our main result, i.e., analytic
results for all master integrals of the banana graph in $d=2$ dimensions in
terms of iterated integrals of modular forms and in terms of elliptic
polylogarithms.  Finally, in section~\ref{sec:conclusions} we draw our
conclusions.  We include additional appendices where we discuss how to obtain
the boundary conditions for the system of differential equations for the banana
graph and where we present a method to decompose an invertible matrix into a
product of a lower and an upper-triangular matrix.

\vspace{1mm}\noindent



\section{The banana graph}
\label{sec:bananareview}

\subsection{Notations and conventions}

\begin{figure}
  \begin{center}
    \begin{tikzpicture}
      \draw[line width=.45mm] (0,0) arc (20:160:2) coordinate (p) node [midway,above]{$m$};
      \draw[line width=.45mm] (0,0) arc (50:130:2.92) node [midway, above]{$m$};
      \draw[line width=.45mm] (0,0) arc (-20:-160:2) node [midway, below]{$m$};
      \draw[line width=.45mm] (0,0) arc (-50:-130:2.92) node [midway, below]{$m$};
      \node(A)[left=1 of p]{};
      \draw[line width=.80mm] (p) ->  (A) node [midway,below]{$p$};
      \draw[line width=.80mm] (0,0) -> (1,0) node [midway,below]{$p$};
      \end{tikzpicture}
  \end{center}
  \caption{The three-loop banana graph.}
  \label{fig:banana}
\end{figure}
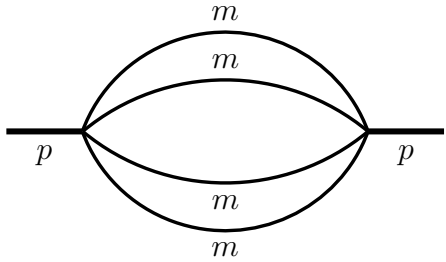

The banana graph 
depicted in fig.~\ref{fig:banana} constitutes one of the simplest families of a three-loop 
Feynman graph. Whenever at least either two propagator
masses or the external invariant vanish, all members of the family can
be expressed in terms of standard multiple polylogarithms (see
e.g.~ref.~\cite{Mastrolia:2002tv}).
If all propagators are massive, new classes of
functions are known to show up~\cite{Bloch:2014qca,Primo:2017ipr}, related to a
specific family of K3 surfaces.  Not much is known in the most general
case 
and in particular no
analytic result is known for the banana family with distinct propagator masses. 

Here we focus on a scenario of intermediate complexity, namely the 
case where all internal
masses are chosen to be different from zero and equal. 
More precisely, let us consider the family of integrals defined by
\beq\bsp
  \label{eq:banana-family} I&_{a_1,\dots,a_9}(p^2,m^2;d)=\\
  & = \int \prod_{i=1}^3 \mathfrak{D}^d \ell_i 
		 \frac{(\ell_3^2)^{a_5}(\ell_1\cdot p)^{a_6}(\ell_2\cdot p)^{a_7}(\ell_3\cdot
  p)^{a_8}(\ell_1\cdot\ell_2)^{a_9}}{[\ell_1^2-m^2]^{a_1}[\ell_2^2-m^2]^{a_2}[(\ell_1-\ell_3)^2-m^2]^{a_3}[(\ell_2-\ell_3-p)^2-m^2]^{a_4}}\,,
\esp\eeq
where the $a_i\ge 0$ are positive integers, and we have introduced
the integration measure 
\begin{equation}
  \label{eqn:intemeasure}
  \int\ddl=\frac{1}{\Gamma\left( 2-\frac{d}{2} \right)}\int\frac{d^d\ell}{i\pi^{d/2}}\,.
\end{equation}
Since all integrals depend on $p^2$ and $m^2$ only, it is convenient to express
their non-trivial functional dependence in terms of the dimensionless ratio 
\begin{equation}
\label{eq:defx}
x = \frac{4m^2}{p^2}\,.
\end{equation}
Furthermore, in what follows we will set $m=1$ for simplicity, since the
dependence on $m$ of the different integrals can be recovered by dimensional
analysis. The integrals may diverge in $d=4$ dimensions. We therefore work in
dimensional regularisation where $d=d_0-2\eps$ with $d_0>0$ a positive integer.
Accordingly, all integrals are interpreted as a Laurent series in the
dimensional regulator~$\eps$.

Let us now focus on the integrals in eq.~\eqref{eq:banana-family}.  Using
integration-by-parts identities~\cite{Chetyrkin:1981qh,Tkachov:1981wb}, we can
express any member of this integral family in terms of four distinct master
integrals.  Moreover, we can use dimensional shift
identities~\cite{Tarasov:1996br,Smirnov:1999wz,Anastasiou:1999bn,Anastasiou:2000mf,Anastasiou:2000kp,Lee:2009dh}
to relate the coefficients of the Laurent expansion of these master integrals
in $d=4-2\eps$ to the corresponding ones in $d=2-2\eps$ dimensions.  Indeed,
since all integrals are IR finite, by lowering the number of dimensions we
improve their UV behaviour. This allows us to choose a finite basis of master
integrals as follows
\eqs{
    \label{eq:fints}
\cI_1(\eps;x) &= (1 + 2 \epsilon ) (1 + 3 \epsilon)I_{1,1,1,1,0,0,0,0,0}(p^2,1;2-2\eps)\, ,\\
\cI_2(\eps;x) &= (1 + 2 \epsilon ) I_{2,1,1,1,0,0,0,0,0}(p^2,1;2-2\eps)\, ,\\
\cI_3(\eps;x) &= I_{2,2,1,1,0,0,0,0,0}(p^2,1;2-2\eps)\,,
}
where $x$ has been  defined in eq.~\eqref{eq:defx}.  The fourth master
integral is the three-loop tadpole with squared propagators, which in our
normalisation evaluates to
\begin{equation}\label{eq:tadpole}
 \cI_0(\eps;x) =  I_{2,2,2,0,0,0,0,0,0}(p^2,1;2-2\eps) = 1\,.
\end{equation}
The main goal of this paper is to present analytic results for the three master
integrals in eq.~\eqref{eq:fints} in $d=2$ dimensions, i.e.~for $\eps=0$. We
stress that this is sufficient to obtain results in $d=4-2\eps$ dimensions up
to terms that are suppressed by powers of $\eps$, as one can easily verify from
the relevant dimensional shift relations~\cite{Tarasov:1996br,Lee:2009dh}.  As
an example, the relation needed to express the master integral $\cI_1(d;x)$ in
terms of the four masters in $d-2$ dimensions reads 
\beq\bsp
\label{eq:D_shift}
\cI_1(d;x) &= c_1(d;x) \cI_1(d-2;x) + c_2(d;x) \cI_2(d-2;x) + c_3(d;x) \cI_3(d-2;x)  \\
&+ c_0(d;x)\cI_0(d-2;x) \,,
\esp\eeq
where the coefficients are
\beq\bsp
    c_1(d;x) &= \frac{1-20x}{3x}+\mathcal{O}(d-4)\,,\\
    c_2(d;x) &= \frac{12-16x(5+7x)}{3x^2}+\mathcal{O}(d-4)\,,\\
    c_3(d;x) &= \frac{8(1-4x)(1+2x(5+2x))}{3x^3}+\mathcal{O}(d-4)\,,\\
    c_4(d;x) &= -2+\mathcal{O}(d-4)\,.\label{eq:coeffdm2}
\esp\eeq
Inspecting eq.~\eqref{eq:D_shift}, we see that expanding the left hand side
around $d=4$ corresponds to expanding the integrals on the right hand side in
two dimensions.  Since all coefficients in eq.~\eqref{eq:coeffdm2} are finite
in this limit, this implies that the finite pieces of the master integrals
expanded close to $d=2$ are sufficient to obtain the finite terms of the
corresponding integrals in $d=4$.  Moreover, since the only divergent integral
on the right hand side is  $\cI _0(d;x)$, it is clear that the poles of the
banana integrals close to $d=4$ arise exclusively from the tadpole integral.

\subsection{The system of differential equations satisfied by the banana family}
It is well known known that master integrals satisfy differential equations in
the external kinematic
variables~\cite{Kotikov:1990kg,Kotikov:1991hm,Kotikov:1991pm,Gehrmann:1999as,Gehrmann:2000zt}.
For the three non-trivial master integrals of the banana graph defined in the
previous subsection, this system of differential equations can be written
as~\cite{Primo:2017ipr}
\begin{equation}
 \label{eq:DEfull}
\partial_x \ivec = \Big[B(x) + \eps D(x)\Big]\ivec + \begin{pmatrix}0\\0\\-\frac{1}{2(4x-1)}\end{pmatrix}\,,
\end{equation}
where the matrices $B(x)$ and $D(x)$ are given by
\begin{align}
B(x) &= 
\begin{pmatrix}
\frac{1}{x} & \frac{4}{x} & 0 \\
\frac{1}{4(1-x)} & \frac{1}{x}+\frac{2}{1-x} & \frac{3}{x}+\frac{3}{1-x} \\
-\frac{1}{8(1-x)} + \frac{1}{8(1-4x)} \,\,&\,\, -\frac{1}{1-x} + \frac{3}{2(1-4x)}
 \,\,&\,\, \frac{1}{x}+\frac{6}{1-4x}-\frac{3}{2(1-x)}
\end{pmatrix}\,,\\
D(x) &= \begin{pmatrix}
\frac{3}{x} & \frac{12}{x} & 0 \\
\frac{1}{1-x} & \frac{2}{x}+\frac{6}{1-x} & \frac{6}{x}+\frac{6}{1-x} \\
-\frac{1}{2(1-x)} + \frac{1}{2(1-4x)} \,\,&\,\, -\frac{3}{1-x}+\frac{9}{2(1-4x)} \,\,&\,\, \frac{1}{x}+\frac{12}{1-4x}-\frac{3}{1-x}
\end{pmatrix}\,.
\end{align}
The inhomogeneity arises from the tadpole master integral in
eq.~\eqref{eq:tadpole}, which does not depend on $x$ and  therefore 
decouples entirely from the system of differential
equations.  Since in this paper we are only concerned
with the value of the integrals in $d=2$ dimensions, we can let $\eps=0$ in
eq.~\eqref{eq:DEfull}, which removes the dependence on the matrix $D(x)$. 
From now on we therefore focus on this simpler system. Introducing the
shorthand $\cI_i(x)\equiv \cI_i(0;x)$ for the master integrals evaluated at
$\eps=0$, the system reads
\begin{equation}\label{eq:DEQ_ep=0}
\partial_x {\begin{pmatrix}\cI_1(x)\\\cI_2(x)\\\cI_3(x)\end{pmatrix}} = B(x)\,{\begin{pmatrix}\cI_1(x)\\\cI_2(x)\\\cI_3(x)\end{pmatrix}} + \begin{pmatrix}0\\0\\-\frac{1}{2(4x-1)}\end{pmatrix}\,.
\end{equation}
Let us sketch how to solve this system using the
method of variation of constants. Assume that we can find a fundamental
solution matrix to eq.~\eqref{eq:DEQ_ep=0}, i.e., a $3\times3$ matrix $\cW(x)$
satisfying the homogeneous equation associated to eq.~\eqref{eq:DEQ_ep=0},
\beq\label{eq:DEQ_hom}
\partial_x\cW(x) = B(x)\,\cW(x)\,.
\eeq
We then see that the vector $(M_1(x),M_2(x),M_3(x))^T$ defined as 
\begin{align}
\label{eq:rotation}
\left(\begin{matrix} \cI_{1}(x) \\ \cI_{2}(x)\\ \cI_{3}(x) \end{matrix} \right)=\cW(x)\left(\begin{matrix} M_{1}(x) \\ M_{2}(x)\\ M_{3}(x) \end{matrix} \right)
\end{align}
satisfies the {inhomogeneous} system of differential equations
\begin{align}
\partial_x \left(\begin{matrix} M_{1}(x) \\ M_{2}(x) \\ M_{3}(x)\end{matrix} \right) =&
\,\,\cW^{-1}(x) \left(\begin{matrix}0 \\ 0 \\ -\frac{1}{2(4x-1)}\end{matrix} \right)\,,
\label{eq:syseps0}
\end{align}
which can now easily be solved by quadrature. We note that $\cW(x)$ is always invertible for generic values of 
$x$ because its columns span the three-dimensional solution space of the homogeneous system 
in eq.~\eqref{eq:DEQ_hom} and are therefore linearly independent. 

Solving the differential equation involves then two steps:
\begin{enumerate}
\item Finding a fundamental solution matrix $\cW(x)$ satisfying the homogeneous differential equation in eq.~\eqref{eq:DEQ_hom}.
\item Solving eq.~\eqref{eq:syseps0} by quadrature. This involves in particular computing integrals over (products of) 
the entries of the fundamental solution matrix.
\end{enumerate}
In the remainder of this section we review how to construct the fundamental
solution matrix $\cW(x)$ in the case of the equal-mass banana graph.  The
entries of $\cW(x)$ are in general transcendental functions, so that the second
step will involve the computation of integrals over transcendental functions.
The main goal of this paper is to show how these integrals can be systematically
performed in terms of iterated integrals over known objects.

\subsection{The fundamental solution matrix}
\label{eq:fundamental_banana}
In general, it can be very complicated to find the fundamental solution matrix of a system of differential equations. In the case of Feynman integrals, 
the fundamental solution matrix can be obtained by studying the maximal cut of the integrals~\cite{Primo:2016ebd,Frellesvig:2017aai,Bosma:2017ens}.
For the equal-mass banana graph, there is an alternative way to solve the homogeneous differential equation in eq.~\eqref{eq:DEQ_hom}~\cite{Bloch:2014qca,Primo:2017ipr}, which we review in the remainder of this section. 

It will be convenient to introduce the following parametrisation of the fundamental solution matrix,
\begin{equation}
\label{eq:homsol}
\cW(x) = \begin{pmatrix}
H_1(x) & J_1(x) & I_1(x) \\
H_2(x) & J_2(x) & I_2(x) \\
H_3(x) & J_3(x) & I_3(x)
\end{pmatrix}\,.
\end{equation}
Next, we convert the linear first-order system of differential equations in eq.~\eqref{eq:DEQ_hom} into a third-order differential equation for the first line of $\cW(x)$,
\beq
\cL^{(3)}_xH_1(x)=\cL^{(3)}_xJ_1(x) = \cL^{(3)}_xI_1(x)=0\,,
\eeq
where $\cL^{(3)}_x$ is the third-order linear differential operator~\cite{Bloch:2014qca,Primo:2017ipr}
\begin{equation}
\label{eq:PFPT}
\cL^{(3)}_x = \partial_x^3+\frac{3(8x-5)}{2(x-1)(4x-1)}\partial_x^2+\frac{4x^2-2x+1}{(x-1)(4x-1)x^2}\partial_x+\frac{1}{x^3(4x-1)}\,.
\end{equation}
This can be achieved in a standard way by using the system of differential equations to re-express 
$H_2(x)$ and $H_3(x)$ in terms of $H_1(x)$ and its derivatives, namely
\beq\bsp
  \label{eq:operators}
 H_2(x) &= \frac{1}{4}(x\,\partial_x-1)\,H_1(x)\\
 H_3(x) &= \frac{1}{12}(x^2(1-x)\,\partial_x^2-x(1+x)\,\partial_x+1)\,H_1(x)\,,
\esp\eeq
yielding in this way a third-order differential equation satisfied by $H_1(x)$.
If a solution for $H_1(x)$ can be found by solving this higher-order equation,
the corresponding solutions for $H_2(x)$ and $H_3(x)$ can be recovered by
differentiating $H_1(x)$ according to eq.~\eqref{eq:operators}.  The same reasoning
can of course be applied to the other two columns of $\cW(x)$, i.e.~to the solutions
$J_i(x)$ and $I_i(x)$.  

In general, solving a third-order differential equation is a formidable task,
and no general algorithm is known for finding the kernel of a generic
third-order linear differential operator. It turns out, however, that the
operator $\cL^{(3)}_x$ is very special, and its solution can be expressed in
terms of the solutions to the following second-order differential operator,
\begin{equation}
 \label{eq:symsquare}
 \cL^{(2)}_x = \partial_x^2 + \frac{8x-5}{2(x-1)(4x-1)}\partial_x-\frac{2x-1}{4x^2(x-1)(4x-1)}\,.
\end{equation}
Specifically, $\cL^{(3)}_x$ is the \emph{symmetric square} of the
operator $\cL^{(2)}_x$~\cite{joyce}, meaning that the three independent solutions of
$\cL^{(3)}_x$ are the products of the two independent solutions of
$\cL^{(2)}_x$. The solutions of $\cL^{(2)}_x$, in turn, can be expressed in
terms of complete elliptic integrals of the first kind. With this insight, one
finds that the three independent homogeneous solutions can be suitably written
as~\cite{joyce,Primo:2017ipr}
\begin{gather}
\begin{aligned}
\label{eq:homsol2}
H_1(x) &= \sqrt{\lambda_+(x)\,\lambda_-(x)}\,{{\K}\big(\lambda_+(x)\big)\,{\K}\big(\lambda_-(x)\big)}\,,\\
J_1(x) &= \sqrt{\lambda_+(x)\,\lambda_-(x)}\,{{\K}\big(\lambda_+(x)\big)\,{\K}\big(1-\lambda_-(x)\big)}\,,\\
I_1(x) &= \sqrt{\lambda_+(x)\,\lambda_-(x)}\,{{\K}\big(1-\lambda_+(x)\big)\,{\K}\big(1-\lambda_-(x)\big)}\,, 
\end{aligned}
\end{gather}
where we defined
\begin{equation}
 \lambda_{\pm}(x) = \frac{4x}{2x+(1-2x)\sqrt{\frac{x-1}{x}}\pm\sqrt{\frac{4x-1}{x}}}\,,
\end{equation}
and $\K$ denotes the complete elliptic integral of the first kind
\beq
{\K}(\lambda) = \int_0^1\frac{dt}{\sqrt{(1-t^2)(1-\lambda t^2)}}\,.
\eeq
By inspecting eq.~\eqref{eq:homsol2}, one might wonder why we have used four
apparently independent building blocks to construct the solutions,
i.e.~${\K}\big(\lambda_+(x)\big)$,  ${\K}\big(\lambda_-(x)\big)$,
${\K}\big(1-\lambda_+(x)\big)$, and ${\K}\big(1-\lambda_-(x)\big)$,
when we stated explicitly that all three solutions can be written as
products of only two independent functions. Indeed, the four functions above
are not independent and the explicit relations among them are non-trivial as
they require to cross the branch cut of 
$\K(x)$
and therefore depend on the prescription we adopt to do so.
Instead, working with an over-complete number of functions has the advantage of
allowing us to choose a compact representation for the solutions, which have
the correct analytic properties. For an
explicit solution in terms of two functions only, see eq.~\eqref{eq:H1_to_Psi1} in the next section.

We have thus obtained the components of the first row of the fundamental
solution matrix in eq.~\eqref{eq:homsol}. The other rows can be obtained from
eq.~\eqref{eq:operators}: they involve derivatives of complete elliptic
integrals of the first kind that are expressible in terms of complete elliptic integrals of
the second kind, 
\beq
{\E}(\lambda) = \int_0^1{dt}\,\sqrt{\frac{1-\lambda t^2}{1-t^2}}\,.
\eeq
The complete set of results for the fundamental solution matrix can be found
for example in ref.~\cite{Primo:2017ipr}. 

The previous discussion makes it clear that, upon inserting the solution for
$\cW(x)$ into eq.~\eqref{eq:syseps0}, the $M_i(x)$ will naturally be expressed
as integrals over products of complete elliptic integrals of the first and second
kind.  This program was carried out in ref.~\cite{Primo:2017ipr}.  It is a
priori not obvious if/how these new classes of iterated integrals can be
expressed in terms of other classes of special functions that have appeared in
Feynman integral computations and/or pure mathematics.  The main goal of this
paper is to show that it is indeed possible to express all master integrals
for the banana family in terms of a known set of special functions: the class
of functions that naturally appear in the solution of the two-loop sunrise
integral family with three equal masses. The connection between the two
families of integrals will be explored in more detail in the next section before we return to our original problem.

\vspace{1mm}\noindent



\section{The geometry associated to the two-loop sunrise graph}
\label{sec:sunrise_review}

\subsection{Relating the equal-mass banana and sunrise graphs} 
The purpose of this subsection is to set the stage for the mathematical objects
that will appear in the analytic result for the master integrals of the
equal-mass banana graph in $d=2$ dimensions presented in section~\ref{sec:bananaforms}.
As anticipated at the end of the previous section, the relevant 
functions will essentially be identical to those appearing in the computation of
the equal-mass two-loop sunrise family
\beq\bsp
\label{eq:sunrise-fam}
S_{a_1,\dots, a_5}&(p^2, m^2; d) \\
&\,=   \int \ddl_1\ddl_2 \frac{(\ell_1\cdot p)^{a_4} (\ell_2\cdot p)^{a_5}}{[\ell_1^2 - m^2]^{a_1} [\ell_2^2 - m^2]^{a_2}  [(\ell_1-\ell_2-p)^2-m^2]^{a_3}}\ ,
\esp\eeq
where the integration measure was defined in eq.~\eqref{eqn:intemeasure}. 

It has been known for a long time~\cite{Sabry,Broadhurst:1987ei} that in the
case where all three propagators are massive, the sunrise integral cannot be
expressed in terms of polylogarithmic functions, but instead requires the introduction
of functions related to elliptic integrals.  

By now we know several analytic representations for the sunrise family, all of
which require the introduction of new classes of functions which generalise
multiple polylogarithms and elliptic integrals.  In the remainder of this section we review the
class of functions relevant to the sunrise graph. As we will see in
section~\ref{sec:bananaforms} below, some of these classes of functions also
appear in the banana graph.

We start with some general facts about the sunrise family. Since all of
these results are in principle well known (cf.,~e.g.,
ref.~\cite{Laporta:2004rb,Remiddi:2016gno}), and all the technical steps are
very similar to the case of the banana family discussed in the previous
section, we will be rather brief and only highlight the main points.  The
equal-mass sunrise family has three master integrals.  One of these master
integrals can be chosen as the tadpole integral $S_{2,2,0,0,0}(p^2, m^2;
2-2\eps)$, which equals one in our normalisation (cf.~eq.~\eqref{eq:tadpole}). 

Following ref.~\cite{Remiddi:2016gno}, we choose the remaining two master
integrals as\footnote{In ref.~\cite{Remiddi:2016gno}, the master integrals
	$\cS_1$ and $\cS_2$ are named $g_6$ and $g_7$ and are defined in
	eq.~(7.7).  The kinematical parameter $u$ in~\cite{Remiddi:2016gno}
equals our parameter $t$, after taking into account that propagators are
Euclidean in the definition (5.1) of the integrals in that reference.}
\beq\bsp
\label{eq:sunrise-basis}
\cS_1 (\eps; t) &= -S_{1,1,1,0,0}(p^2, m^2; 2-2\eps) \ ,\\
\cS_2 (\eps; t) &= -\left[ \frac{1}{3}(t^2-6t+21) -12\eps(t-1)\right]
    S_{1,1,1,0,0}(p^2, m^2; 2-2\eps)\\
    &\quad- 2 (t-1)(t-9) S_{2,1,1,0,0}(p^2, m^2; 2-2\eps),
\esp\eeq
where we encode the kinematics in the dimensionless variable $t = p^2/m^2$.
Note that the variable $t$ should not be confused with the variable $x$ defined  in eq.~\eqref{eq:defx} 
for the banana graph: the two quantities are not trivially related.
The precise relation between the quantity $x$ for the banana family and the kinematical variable $t$ defined here will be discussed below. 

Just like for the banana family, we will put $m=1$ in the following, as its dependence
can be restored later on by simple dimensional analysis.  The master integrals
in eq.~\eqref{eq:sunrise-basis} satisfy the following system of differential
equations~\cite{Remiddi:2016gno},
\beq
\bsp
\label{eq:sunrise-deq}
\partial_t \begin{pmatrix} \cS_1 (\eps; t)  \\ \cS_2 (\eps; t) \end{pmatrix} \,=\, (\tilde{B}(t) - 2 \eps \tilde{D}(t)) \begin{pmatrix} \cS_1 (\eps; t)   \\ \cS_2 (\eps; t)  \end{pmatrix}  + \begin{pmatrix}0 \\ 1 \end{pmatrix}\ ,
\esp
\eeq
where $\tilde{B}(t)$ and $\tilde{D}(t)$ are $2 \times 2$ matrices which are
{independent} of $\epsilon$, while the inhomogeneous term comes from the
tadpole master integral which decouples from the system of differential
equations.  For simplicity, in the following we focus on the sunrise family in
$d=2$ dimensions. The master integrals $\cS_i(\eps; t) = \cS_i(t)+\cO(\eps)$
are finite in two dimensions, so we can let $\eps=0$ in
eq.~\eqref{eq:sunrise-deq} and ignore the contribution from $\tilde{D}(t)$. The
matrix $\tilde{B}(t)$ is given by ref.~\cite{Remiddi:2016gno}. Adapted to our conventions, it reads,
\beq\bsp
\label{eq:sunrise-deq-matrices}
\tilde{B}(t)\,&=\, \frac{1}{6 \,t\, (t-1) (t-9)}
\begin{pmatrix}
3(3+14t-t^2) & -9 \\
(t+3)(3+75 t -15 t^2 + t^3)  & -3(3+14t -t^2)
\end{pmatrix}\,.
\esp\eeq

We first have to solve the homogenous equation associated to
eq.~\eqref{eq:sunrise-deq}, i.e., we need to find a $2\times2$ matrix
$\cW_S(t)$ that satisfies $\partial_t\cW_S(t) = \tilde{B}(t)\cW_S(t)$. The solution to
the inhomogeneous equation for $\eps=0$ in eq.~\eqref{eq:sunrise-deq} is then
obtained by defining the new basis $(\cS_1(t),\cS_2(t))^T =
\cW_S(t)(\cT_1(t),\cT_2(t))^T$ which fulfils the simpler {inhomogeneous}
differential equation,
\beq\label{eq:sunrise_T}
\partial_t\left(\begin{matrix}\cT_1(t)\\ \cT_2(t)\end{matrix}\right) = \cW_S(t)^{-1}\begin{pmatrix} 0 \\ 1 \end{pmatrix}\,.
\eeq
The $2 \times 2$ system satisfied by $\cW_S(t)$ is equivalent to
a linear second-order differential equation for the functions in the first row 
of $\cW_S(t)$~\cite{Laporta:2004rb}:
\begin{equation}\label{eq:DSunrisePT}
  \cL_t^{(2)}=\partial_t^2 + \Big(\frac{1}{t-9}+\frac{1}{t-1}+\frac{1}{t}\Big)\partial_t+\Big(\frac{1}{12(t-9)}+\frac{1}{4(t-1)}-\frac{1}{3t}\Big).
  \end{equation}
We choose its kernel to be spanned\footnote{In terms of the solutions in
	ref.~\cite{Remiddi:2016gno} one finds that for the region $0<t<1$ we can relate the solutions as  
	$\Psi_1(t)=2 I_1^{(0,1)}(t)$ and $\Psi_2(t)=2iJ_1^{(0,1)}(t)$, where the
	integrals on the right-hand side are defined in eq.~(D.11) in ref.~\cite{Remiddi:2016gno}.} by the functions $\Psi_1$ and $\Psi_2$:
\beq\bsp
  \label{eq:psi1_def}
  \Psi_1(t) & = \frac{4}{[(3-\sqrt{t})(1+\sqrt{t})^3]^{1/2}}\,{\K}\left(\frac{t_{14}(t)t_{23}(t)}{t_{13}(t)t_{24}(t)}\right)\,,\\
  \Psi_2(t) & = \frac{4 i}{[(3-\sqrt{t})(1+\sqrt{t})^3]^{1/2}}\,{\K}\left(\frac{t_{12}(t)t_{34}(t)}{t_{13}(t)t_{24}(t)}\right)\,,
\esp\eeq
with $t_{ij}(t) = t_i(t)-t_j(t)$ and 
\beq\label{eq:SR_t_i_def}
t_1(t) = -4\,,\quad t_2(t) = -(1+\sqrt{t})^2\,,\quad t_3(t) = -(1-\sqrt{t})^2\,, \quad t_4(t)=0\,.
\eeq
The period matrix for the sunrise differential equation is then,
\begin{equation}\label{eq:sunrise_wronskian}
    \mathcal{W}_S(t) = \begin{pmatrix}
        \Psi_1(t) & \Psi_2(t) \\
        \mathcal{D}_t\Psi_1(t) & \mathcal{D}_t\Psi_1(t)\\
    \end{pmatrix}\,,
\end{equation}
with $\mathcal{D}_t =
\tfrac{1}{3}(3+14t-t^2)-\tfrac{2}{3}(t-9)(t-1)\partial_t$.  Note that
$\Psi_1(t)$ and $\Psi_2(t)$ are naturally related to the maximal cut of the
integral $S_{1,1,1,0,0}(p, m^2; 2)$~\cite{Laporta:2004rb}. It turns out that
the second-order differential operator $\cL_t^{(2)}$ in
eq.~\eqref{eq:DSunrisePT} is closely related to the second-order operator
$\cL_x^{(2)}$ for the banana graph provided in eq.~\eqref{eq:symsquare}.
Indeed, relating the kinematical variables for the sunrise and the banana graph
via
\begin{equation}\label{eq:x_to_t}
  x(t)=\frac{-4\,t}{(t-1)(t-9)}\,,
\end{equation}
one finds
\begin{equation}\label{eq:alternativeOperator}
	\cL_x^{(2)}
	=\tilde{\cL}_t^{(2)}=\partial_t^2 + \Big(\frac{1}{t-9}+\frac{1}{t-1}\Big)\partial_t+\Big(\frac{1}{36(t-9)}-\frac{1}{4(t-1)}+\frac{1}{4 t^2}+\frac{2}{9t}\Big).
  \end{equation}
This is not quite the same operator as in eq.~\eqref{eq:DSunrisePT}. However, one can verify that 
\begin{align}
  \label{eq:tildeLt}
  \tilde{\cL}_t^{(2)}\sqrt{t}\Psi_1(t) = \tilde{\cL}_t^{(2)}\sqrt{t}\Psi_2(t) =0\,,
\end{align}
that is, the solutions to the two differential operators differ by a square root of $t$. 

In section~\ref{eq:fundamental_banana} we stated that the third-order differential operator $\cL_x^{(3)}$ in
eq.~\eqref{eq:PFPT} is the symmetric square of $\cL_x^{(2)}$. Correspondingly, the
solutions of  $\cL_x^{(3)}$ are sums of products of the functions in
eq.~\eqref{eq:psi1_def} with an additional factor of $(\sqrt{t})^2=t$ which can
be precisely traced back to eq.~\eqref{eq:tildeLt}. 
In particular, it is straightforward to check that the
functions in eq.~\eqref{eq:homsol} can be cast in the following alternative
form which makes manifest the connection between the fundamental solution
matrix for the banana graph, $\cW(x)$, and the one for the sunrise, $\cW_S(t)$, namely
\beq\bsp\label{eq:H1_to_Psi1}
H_1(x(t)) &\,= -\frac{1}{3}\,t\,\Psi_1(t)^2\,,\\
J_1(x(t)) &\,=\frac{i}{3}\,t\,\Psi_1(t)\,(\Psi_1(t)+\Psi_2(t))\,,\\
I_1(x(t)) &\,=\frac{1}{3}\,t\,(\Psi_1(t)+\Psi_2(t))\,(\Psi_1(t)+3\Psi_2(t))\,.
\esp\eeq
We see that, as expected, the solutions of $\cL_x^{(3)}$ are sums of products
of the solutions of $\cL_x^{(2)}$ with an additional prefactor of $t$. 

Equation~\eqref{eq:H1_to_Psi1} is our first hint that the function spaces of
the sunrise and banana families in $d=2$ dimensions are closely related.  Since
the two-loop sunrise graph can be expressed in terms of elliptic
polylogarithms~\cite{Bloch:2013tra,Adams:2013kgc,Adams:2013nia,Adams:2014vja,Adams:2015gva,Adams:2015ydq,Broedel:2017siw}
and iterated integrals of modular forms~\cite{Adams:2017ejb,Broedel:2018iwv},
it is tantalising to investigate whether the same class of functions describes the
banana family in $d=2$ dimensions as well. This was already hinted at in
ref.~\cite{Bloch:2014qca}, where it was argued that the three-loop equal-mass
banana graph is an elliptic trilogarithm and closely related to the same
congruence subgroup relevant to the two-loop equal-mass sunrise graph. In the
remainder of this paper we make this connection concrete, and we present
analytic results for the equal-mass banana graph in $d=2$ dimensions in terms
of the same class of functions as for the two-loop equal-mass sunrise graph.

\subsection{The elliptic curve associated to the sunrise graph}

Since the goal of this paper is to show that the equal-mass sunrise and banana
graphs can be expressed in terms of the same class of functions, let us review
in the remainder of this section the geometric objects and functions that
appear in the computation of the two-loop equal-mass sunrise graph.

In the previous section we have seen that the homogeneous solutions of the
second order differential equation satisfied by the two-loop equal-mass sunrise
graph can be expressed in terms complete elliptic integrals of the first kind,
cf.~eq.~\eqref{eq:psi1_def}. 
The appearance of complete elliptic integrals is closely related to
the presence of an elliptic curve in the geometry associated to the problem.
Loosely speaking, an elliptic curve can be defined as the set of points $(x,y)$
that solve the polynomial equation $y^2=(x-a_1)\cdots (x-a_4)$, where the $a_i$
are complex numbers that are constants with respect to $(x,y)$. Instead of
characterising an elliptic curve by the roots $a_i$ of the polynomial equation,
we can also characterise it by its two \emph{periods}, defined by
\beq\label{eq:periods_def}
\omega_1 = 2\,\K(\lambda) \textrm{~~and~~}\omega_2 = 2i\,\K(1-\lambda)\,, 
\textrm{~~with~~}\lambda=\frac{(a_1-a_4)(a_2-a_3)}{(a_1-a_3)(a_2-a_4)}\,.
\eeq

The periods are not uniquely defined, but we could replace them by any integer
linear combination of the $\omega_1$ and $\omega_2$ chosen above.  More
precisely, the periods are only defined modulo $\SL(2,\mathbb{Z})$
transformations, which act on the two periods as follows
\beq
\left(\begin{matrix}\omega_2\\ \omega_1\end{matrix}\right)  \to  \left(\begin{matrix}a & b \\ c & d\end{matrix}\right) \left(\begin{matrix}\omega_2\\ \omega_1\end{matrix}\right)\,,\qquad \left(\begin{matrix}a & b \\ c & d\end{matrix}\right) \in\SL(2,\mathbb{Z})\,.
\eeq
Such transformations are called \emph{modular transformations}.  The geometry
is also left unchanged by a rescaling, and so only the ratio of the two periods
carries relevant information
\beq
\tau = \frac{\omega_2}{\omega_1} = i \frac{\K(1-\lambda)}{\K(\lambda)}\,,
\eeq
where it is customary to refer to $\tau$ as the \emph{modular parameter} of the
elliptic curve.  Modular transformations act on $\tau$ via M\"obius
transformations,
\beq
\tau \to \frac{a\tau+b}{c\tau+d}\,,\qquad \left(\begin{matrix}a & b \\ c & d\end{matrix}\right) \in\SL(2,\mathbb{Z})\,.
\eeq
Note that it is always possible to choose $\tau$ to lie in the complex upper
half-plane $\mathbb{H}=\{\tau\in\mathbb{C}\,|\,\textrm{Im
}\tau>0\}$.

In many situations one is not interested in modular transformations associated
with the full group $\SL(2,\mathbb{Z})$, but only a subgroup $\Gamma\subset
\SL(2,\mathbb{Z})$ is relevant. In particular, in many applications in
mathematics and physics the various \emph{congruence subgroups of level $N$}
play a prominent role,
\beq\bsp\label{eq:congruence_subgroups}
\Gamma_0(N) &\, = \left\{\left(\begin{smallmatrix}a&b\\ c&d\end{smallmatrix}\right)\in\SL(2,\mathbb{Z}) \big|\, c \equiv 0 \mymod N\right\}\,,\\
\Gamma_1(N) &\, = \left\{\left(\begin{smallmatrix}a&b\\ c&d\end{smallmatrix}\right)\in\SL(2,\mathbb{Z}) \big|\, a,d\equiv 1\mymod N\textrm{~and~}c \equiv 0 \mymod N\right\}\,,\\
\Gamma(N) &\, = \left\{\left(\begin{smallmatrix}a&b\\ c&d\end{smallmatrix}\right)\in\SL(2,\mathbb{Z}) \big|\, a,d\equiv 1\mymod N\textrm{~and~}b,c\equiv 0 \mymod N\right\}\,.
\esp\eeq

Let us now discuss how a family of elliptic curves arises from the sunrise
graph.  We see from eq.~\eqref{eq:periods_def} that the periods of an elliptic
curve can be expressed in terms of complete elliptic integrals of the first
kind.  The same is true for the functions $\Psi_1(t)$ and $\Psi_2(t)$, which
define two independent periods of a family of elliptic curves parametrised by
the parameter $t$. The polynomial equation describing a member of this family
is $y^2=(x-t_1(t))\ldots(x-t_4(t))$, where the $t_i(t)$ were defined in
eq.~\eqref{eq:SR_t_i_def}. A member of this family can be defined equivalently
by specifying the value of $t$ or of the modular parameter $\tau$,
\begin{equation}
  \label{eq:psi3def}
  \tau=\frac{\Psi_2(t)}{\Psi_1(t)}\,.
\end{equation}
It is possible to invert eq.~\eqref{eq:psi3def} and express $t$ as a function of $\tau$~\cite{Maier},
\begin{equation}
  \label{eq:t16}
  t(\tau)=9\frac{\eta(\tau)^4\eta(6\tau)^8}{\eta(2\tau)^8\eta(3\tau)^4}\,,
\end{equation}
where $\eta(\tau)$ denotes the Dedekind $\eta$-function,
\begin{equation}
  \eta(\tau)=q^{1/24}\prod_{n=1}^\infty(1-q^n),\quad q=e^{2\pi i \tau}\,.
\end{equation}
The function $t(\tau)$ is invariant under modular transformations for $\Gamma_1(6)$,
\begin{align}
  t\left(\frac{a\tau + b}{c\tau +d}\right) = t(\tau)\,,\qquad \left(\begin{matrix}a & b \\ c & d\end{matrix}\right) \in\Gamma_1(6)\,.
\end{align}
Therefore, the family of elliptic curves associated to
the sunrise graph is tightly related to the congruence subgroup
$\Gamma_1(6)$~\cite{Bloch:2013tra,Adams:2017ejb}.\footnote{Depending on whether one
starts from the elliptic curve obtained from the Feynman parameter integral or
the maximal cut, one may instead find $\Gamma_1(12)$, see
ref.~\cite{Adams:2017ejb} for a detailed discussion.}

In general, we need to consider not only functions that are invariant under
\mbox{$\Gamma\subset \SL(2,\mathbb{Z})$}, but also functions with non-trivial
transformation behaviour.  A \emph{modular form of weight $n$ for $\Gamma$} is
a holomorphic function $f$ which transforms covariantly under modular
transformations for the group $\Gamma$,
\begin{align}
  f\left(\frac{a\tau + b}{c\tau +d}\right) = (c\tau+d)^n f(\tau)\,,\qquad \left(\begin{matrix}a & b \\ c & d\end{matrix}\right) \in\Gamma\,,
\end{align}
subject to some regularity conditions which we can ignore at this point. It is
easy to see that modular forms define an algebra: the product of two
modular forms of weights $n_1$ and $n_2$ is a modular form of weight $n_1+n_2$.
If we denote by $\cM_n(\Gamma)$ the vector space of modular forms of weight $n$
for $\Gamma$, then $\cM_n(\Gamma)$  is always finite-dimensional. It is
possible to construct bases for $\cM_n(\Gamma)$ in a completely algorithmic
way. Here we only discuss the case $\Gamma = \Gamma_1(6)$, which is relevant to
the computation of the equal-mass sunrise and banana graphs. The basis
described below was introduced in ref.~\cite{Broedel:2018rwm}. 

We start by noting that the function
\beq
f_{1,0}(\tau) = \Psi_1(t(\tau))
\eeq
is a modular form of weight one for
$\Gamma_1(6)$~\cite{Adams:2017ejb,Broedel:2018rwm}. Since modular forms form an
algebra, it is clear that $(f_{1,0}(\tau))^n$ will define a modular form of
weight $n$. Moreover, since $t(\tau)$ in eq.~\eqref{eq:t16} is invariant under
$\Gamma_1(6)$, multiplying powers of $f_{1,0}(\tau)$ by any (rational) function
of $t(\tau)$ will not change the behaviour under modular transformations for
$\Gamma_1(6)$. The requirement that modular forms be holomorphic everywhere
restricts these rational functions to be polynomials. The maximal power of this
polynomial can be constrained by analysing the behaviour of $\Psi_1(t)$ for
large values of $t$ (for details see ref.~\cite{Broedel:2018rwm}). With these
considerations, one finds that a basis of $\cM_n(\Gamma_1(6))$ is given by the
functions~\cite{Broedel:2018rwm}
\beq\label{eq:Gamma_1(6)_basis}
f_{n,p}(\tau) = \Psi_1(t(\tau))^n\,t(\tau)^p\,,\quad 0\le p\le n\,.
\eeq
Note that this definition extends to modular forms of weight zero,
$f_{0,0}(\tau)=1$.  The advantage of this basis in the context of the sunrise
and banana graphs will be discussed in the remainder of this section.

\subsection{A class of iterated integrals of modular forms for $\Gamma_1(6)$}
\label{ssec:itintmodformsg16}
After this excursion into the geometry associated to the sunrise graph, let us
now review what it can teach us about the functions the sunrise graph
evaluates to.  It is known that the two-loop equal-mass sunrise integral can be
expressed in terms of iterated integrals of modular forms for
$\Gamma_1(6)$~\cite{Adams:2017ejb}.  In this section we give a short review of
these functions with a special focus on the case of $\Gamma_1(6)$. 

If $f_{i_a}(\tau)$ are modular forms of weight $n_{i_a}$ for a congruence
subgroup $\Gamma$, we define the iterated
integrals~\cite{ManinModular,Brown:mmv}
\beq\label{eq:IMF_gen_def}
\mathcal{I}(f_{i_1},\ldots,f_{i_k};\tau) = \int_{i\infty}^{\tau}d\tau'\,f_{i_1}(\tau')\,\mathcal{I}(f_{i_2},\ldots,f_{i_k};\tau')\,.
\eeq
In general these integrals may diverge, but the divergences can be regulated in a
standard way~\cite{Brown:mmv} (see also ref.~\cite{Adams:2017ejb} for a
pedagogical introduction). Moreover, these integrals satisfy all the properties
of iterated integrals. In particular they form a shuffle algebra.  We define
the length of $\mathcal{I}(f_{i_1},\ldots,f_{i_k};\tau)$ as $k$. 

Let us now discuss how we can associate a concept of transcendental weight to the functions
$\mathcal{I}(f_{i_1},\ldots,f_{i_k};\tau)$. If $\Gamma$ is a congruence subgroup
of level $N$, then modular forms for $\Gamma$ are invariant under translations
by $N$, $f_{i_a}(\tau+N) = f_{i_a}(\tau)$. Hence, $f_{i_a}(\tau)$ admits a
Fourier series of the form
\beq
\label{eq:q-exp}
f_{i_a}(\tau) = \sum_{n=0}^\infty a_n\,q_N^n\,,\quad q_N=e^{2\pi i\tau/N}\,.
\eeq
It is always possible to choose a basis such that the Fourier coefficients are
rational multiples of $\pi^{n_{i_a}}$. With this normalisation, we define the
transcendental weight\footnote{Note that the transcendental weight of the
	iterated integrals is distinct from the weight of a modular form under
modular transformations. In particular, the iterated integrals will in general
not be modular forms.} of $\mathcal{I}(f_{i_1},\ldots,f_{i_k};\tau)$ to be
$\sum_{a=1}^kn_{i_a}$. The rationale behind this definition will become clear
in the next section.

In the case $\Gamma=\Gamma_1(6)$, we can work with the explicit basis of
modular forms in eq.~\eqref{eq:Gamma_1(6)_basis} and we define
\beq\label{eq:IMF_def}
\IMF{n_1&\ldots & n_k\\ p_1 &\ldots& p_k}{\tau} = \mathcal{I}(f_{n_1,p_1},\ldots,f_{n_k,p_k};\tau)\,.
\eeq
It is easy to check that the modular forms $f_{n,p}(\tau)$ are normalised such
that their Fourier coefficients are proportional to $\pi^n$. Hence,
$\IMF{n_1&\ldots & n_k\\ p_1 &\ldots& p_k}{\tau}$ has length $k$ and weight
$\sum_{a=1}^kn_a$.

The iterated integrals $\IMF{n_1&\ldots & n_k\\ p_1 &\ldots& p_k}{\tau}$ have
an important property: they allow for an alternative description in terms of
iterated integrals over products of complete elliptic integrals, similar to
those that have appeared in refs.~\cite{Remiddi:2016gno,Primo:2017ipr} in the
context of the sunrise and banana graphs. The basic idea is the following: we
see from eq.~\eqref{eq:Gamma_1(6)_basis} that if we change variables from
$\tau$ to $t$ using eq.~\eqref{eq:psi3def}, then  $f_{n,p}(\tau)$ is
proportional to $\Psi_1(t)^n$. The Jacobian of the change of variables is given
by
\beq\label{eq:dtau_to_dt}
d\tau = -\frac{6\pi i\,dt}{t(t-1)(t-9)\,\Psi_1(t)^2}\,,
\eeq
where we used the fact that
\begin{equation}\bsp\label{eq:Wronskiansunrise}
    \det\mathcal{W}_S(t)
    = -\frac{2}{3}t(t-9)(t-1)\left[\Psi_1(t)\,\partial_t\Psi_2(t)-\Psi_2(t)\,\partial_t\Psi_1(t)\right]
    = 4\pi i\,.
\esp\end{equation}
Hence, the integration kernels that define the
iterated integrals can be cast in the form
\beq
d\tau\,f_{n,p}(\tau) = -\frac{6 \pi i\,dt\,t^{p-1}}{(t-1)(t-9)}\,\Psi_1(t)^{n-2}\,,
\eeq
and in this way 
we obtain an alternative description of the iterated integrals for $\Gamma_1(6)$ 
as iterated integrals over products of complete elliptic integrals,
\beq
\IMF{n_1&\ldots & n_k\\ p_1 &\ldots& p_k}{\tau} = -6\pi i\int_0^{t(\tau)}\frac{dt'\,t'^{p_1-1}}{(t'-1)(t'-9)}\,\Psi_1(t')^{n_1-2}\,\IMF{n_2&\ldots & n_k\\ p_2 &\ldots& p_k}{\tau'(t')}\,.
\eeq
We see that the basis of modular forms in eq.~\eqref{eq:Gamma_1(6)_basis} and
the iterated integrals in eq.~\eqref{eq:IMF_def} allow us to easily switch
between the two representations in terms of modular forms or products of
complete elliptic integrals. This observation will be the key to expressing the
master integrals for the banana family as iterated integrals of modular forms
for $\Gamma_1(6)$.  Before we do this, we find it instructive to review the
same procedure in the context of the master integrals for the sunrise family.

\subsection{The sunrise integral and modular forms for $\Gamma_1(6)$}
\label{ssec:sunriseG16}
To see how the two-loop sunrise integral can be expressed in terms of iterated integrals
of modular forms, we start from the differential equation in
eq.~\eqref{eq:sunrise_T}, which we rewrite as
\beq\label{eq:sunrise_T_2}
\partial_t\left(\begin{matrix}\cT_1(t)\\ \cT_2(t)\end{matrix}\right) =\frac{1}{4 \pi i} \begin{pmatrix} -\Psi_2(t) \\ \phantom{-}\Psi_1(t) \end{pmatrix}\,.
\eeq
We change variables from $t$ to the modular parameter $\tau$ using
eq.~\eqref{eq:t16}. The Jacobian of the change of variables can easily be read
of from eq.~\eqref{eq:dtau_to_dt}. We find
\beq
\label{eq:dtaudt}
\partial_{\tau} = -\frac{1}{6\pi i}\,t(\tau)(t(\tau)-1)(t(\tau)-9)\,\Psi_1(t(\tau))^2\,\partial_t\,,
\eeq
and so eq.~\eqref{eq:sunrise_T_2} becomes  
\beq\bsp
\label{eq:sunrise_T_2_tau}
\partial_{\tau}\left(\begin{matrix}\cT_1(t(\tau))\\ \cT_2(t(\tau))\end{matrix}\right) &\,= \frac{1}{24\pi^2}\,t(\tau)(t(\tau)-1)(t(\tau)-9)\,\Psi_1(t(\tau))^2\, \begin{pmatrix} -\Psi_2(t) \\ \phantom{-}\Psi_1(t) \end{pmatrix}\\
&\,= \frac{1}{24\pi^2}\,\left(f_{3,3}(\tau)-10f_{3,2}(\tau)+9f_{3,1}(\tau)\right)\,\begin{pmatrix} -\tau\\1\end{pmatrix}\\
&\,= \frac{1}{24\pi^2}\,\left(f_{3,3}(\tau)-10f_{3,2}(\tau)+9f_{3,1}(\tau)\right)\,\begin{pmatrix} -\IMF{0\\0}{\tau}\\1\end{pmatrix}\, , 
\esp\eeq
where we used the fact that
\beq
\label{eq:tau-integral}
\tau = \int_{i\infty}^\tau d\tau^\prime f_{0,0}(\tau^\prime) = \IMF{0\\0}{\tau}\, .
\eeq
We can choose as initial condition the point $t=0$, which corresponds to
$\tau\to i\infty$. Translating the results of ref.~\cite{Remiddi:2016gno} to
our conventions we find
\beq\bsp
\mathcal{T}_1(t) &\,= {\rm Cl}_2(\pi/3) + \cO(t)\,,\\
\mathcal{T}_2(t) &\,= 0 + \cO(t)\,.
\esp\eeq
Here ${\rm Cl}_2(x)$ denotes the Clausen function,
\beq
{\rm Cl}_2(x)\,=\, \frac{i}{2}(\text{Li}_2(e^{-i x})-\text{Li}_2(e^{i x}))\ .
\eeq
We then find the following result for $\mathcal{T}_i$,
\beq\bsp
\label{eq:finalresult-sunrise}
\mathcal{T}_1(t(\tau)) &\,= \frac{{\rm Cl}_2(\pi/3)}{2\pi} -\frac{1}{24\pi^2 }\,\left[\IMF{3 & 0\\ 3&0}{\tau} - 10\, \IMF{3 & 0\\ 2&0}{\tau} + 9\,\IMF{3 & 0\\ 1&0}{\tau}\right]\,,\\
\mathcal{T}_2(t(\tau)) &\,= \frac{1}{24\pi^2}\,\left[\IMF{3 \\ 3}{\tau} - 10\, \IMF{3 \\ 2}{\tau} + 9\,\IMF{3\\ 1}{\tau}\right]\,.
\esp\eeq

Let us make a comment about the form of the result for the sunrise graph in
eq.~\eqref{eq:finalresult-sunrise}.  It is easy to see that the result in
eq.~\eqref{eq:finalresult-sunrise} is a linear combination of functions of
uniform weight one, where the weight of the iterated integrals of modular forms
was defined earlier, and the weight of Clausen function and $\pi$ is defined in
the usual way. This fact was first observed in ref.~\cite{Broedel:2018qkq}.

\vspace{1mm}\noindent



\section{Analytic results for the equal-mass banana graph}
\label{sec:bananaforms}

After a brief detour through the sunrise integral family, in this section we return to the banana family and present the main results of this paper.
We derive fully analytic results for all master integrals for the equal-mass banana graphs in $d=2$ dimensions. In order to achieve this, we proceed in exactly the same way as for the sunrise graph in the previous section: we start by showing how we can relate the fundamental solution matrix of the system of differential equations satisfied by the master integrals for the banana family, eq.~\eqref{eq:DEQ_ep=0}, to modular forms for $\Gamma_1(6)$. In particular, we express the results for all master integrals in terms of the iterated integrals of modular forms for $\Gamma_1(6)$ defined earlier in eq.~\eqref{eq:IMF_def}.

After representing the equal-mass banana integral in terms of modular forms for $\Gamma_1(6)$, we carry on with the main theme of this paper and ask whether the banana integral can also be recast in terms of other functions used in the past to represent the sunrise integral~\cite{Adams:2017ejb}. We then show the result for the banana integral in terms of iterated integrals of Eisenstein series of level six and elliptic multiple polylogarithms (eMPLs).

\subsection{The equal-mass banana graph and modular forms for $\Gamma_1(6)$}
We start from eq.~\eqref{eq:H1_to_Psi1}, which relates the entry $H_1(x)$ in the fundamental solution matrix $\cW(x)$ to the maximal cut of the equal-mass sunrise graph in $d=2$ dimensions. Comparing eqs.~\eqref{eq:H1_to_Psi1} and~\eqref{eq:Gamma_1(6)_basis}, we immediately see that 
\beq\bsp\label{eq:HJI_to_tau}
H_1(x(\tau)) &\,= -\frac{1}{3}\,f_{2,1}(\tau)\,,\\
J_1(x(\tau)) &\,=\frac{i}{3}\,f_{2,1}(\tau)\,(1+\tau)\,,\\
I_1(x(\tau)) &\,=\frac{1}{3}\,f_{2,1}(\tau)\,(1+\tau)\,(1+3\tau)\,.
\esp\eeq
where $x(\tau)$ is obtained by composing eq.~\eqref{eq:x_to_t} with
eq.~\eqref{eq:t16}, and can be written as~\cite{verrill1996,Bloch:2014qca}
\begin{equation}
  \label{eqn:relxtau}
  x(\tau) = -4\left(\frac{\eta(2\tau)\eta(6\tau)}{\eta(\tau)\eta(3\tau)}\right)^6 \,.
\end{equation}
We see that after changing variables from $x$ to $\tau$, $H_1(x(\tau))$ is a
modular form of weight two for $\Gamma_1(6)$, while $J_1(x(\tau))$ and
$I_1(x(\tau))$ are modular forms multiplied by a polynomial in $\tau$.  The
other entries in $\cW(x)$ also involve derivatives of $\Psi_1(t)$ and
$\Psi_2(t)$, and so they cannot be expressed in terms of $f_{2,1}(\tau)$ and
$\tau$ alone. 

In a next step, let us rewrite the  differential equations for the three master
integrals of the banana graph, eq.~\eqref{eq:syseps0}, in terms of the modular
parameter $\tau$ and express them in the language of modular forms for
$\Gamma_1(6)$.  In order to do so, we need to include the Jacobian from the
change of variables from $x$ to $\tau$, whose value is easily obtained by
combining eq.~\eqref{eq:x_to_t} with eq.~\eqref{eq:dtau_to_dt}. We find,
\beq
\partial_\tau = -\frac{2\,t(\tau)(t(\tau)^2-9)}{3\pi i\,(t(\tau)-1)(t(\tau)-9)}\,\Psi_1(t(\tau))^2\,\partial_x\,.
\eeq
The desired differential equation immediately follows upon expressing the
elements of the fundamental solution matrix $\cW(x)$ in terms of the basis of
modular forms for $\Gamma_1(6)$ in eq.~\eqref{eq:Gamma_1(6)_basis} and
inverting it. Note that the expression for $\cW(x)^{-1}$ involves derivatives
of $\Psi_1(t)$ and $\Psi_2(t)$, but we find that the dependence on the
derivatives drops out once eq.~\eqref{eq:Wronskiansunrise} is imposed.
Combining everything, we find 
\begin{align}
\notag
 \partial_\tau& \left(\begin{matrix} M_{1}(x(\tau)) \\ M_{2}(x(\tau)) \\ M_{3}(x(\tau))\end{matrix} \right) =
  \frac{f_{4,4}(\tau)-10f_{4,3}(\tau)+90f_{4,1}(\tau)-81f_{4,0}(\tau)}{18 i \pi^3}\left(\begin{matrix} 
 3 (1+\tau)^2\\
 -2i(2+3\tau)\\
-1
 \end{matrix} \right)\\
  \label{eq:7.2tau}&\,=  \frac{f_{4,4}(\tau)-10f_{4,3}(\tau)+90f_{4,1}(\tau)-81f_{4,0}(\tau)}{18 i \pi^3}\left(\begin{matrix} 
 3 (1+\IMF{0\\0}{\tau})^2\\
 -2i(2+3\IMF{0\\0}{\tau})\\
-1
 \end{matrix} \right)\, .
\end{align}
The above differential equation can be solved by quadrature using the iterated
integrals defined in eq.~\eqref{eq:IMF_def}.  The initial condition of the
differential equation can be obtained by analysing the behaviour of the master
integrals $\cI_i(x)$ in eq.~\eqref{eq:fints} as $x\to0$, which corresponds to
$\tau\to i\infty$, the lower integration limit of the iterated integrals in
eq.~\eqref{eq:IMF_gen_def}. We find,
\begin{align}
    \label{eq:mba}
    \begin{split}
        \cI_1(x)&=x\log^3(-x/4)-4x\zeta_3 + \cO(x^2)\,,\\
        \cI_2(x)&=\tfrac{3}{4}x\log^2(-x/4) + \cO(x^2)\,,\\
        \cI_3(x)&=\tfrac{1}{2}x\log(-x/4) + \cO(x^2)\,.
    \end{split}
\end{align}
The derivation of eq.~\eqref{eq:mba} is straightforward but technical. We refer
to appendix~\ref{app:initial_condition} for the details.  Putting everything
together, we find the following result for the master integrals of the
equal-mass banana family in $d=2$ dimensions, 
\begin{align}
\nonumber
    M_{1}(x(\tau)) &= -\frac{4\zeta_3}{\pi^2}-
   \frac{i}{6 \pi^3}\big[81 \IMF{4\\0}{\tau} - 90 \IMF{4\\1}{\tau} + 10 \IMF{4\\3}{\tau} - 
  \IMF{4\\4}{\tau} \\
\nonumber  &\qquad\qquad+ 162\IMF{4&0\\0&0}{\tau}  - 
  180 \IMF{4&0\\1&0}{\tau} + 20\IMF{4&0\\3&0}{\tau} - 
  2 \IMF{4&0\\4&0}{\tau}\\
\nonumber&\qquad\qquad+ 162 \IMF{4&0&0\\0&0&0}{\tau} - 
  180 \IMF{4&0&0\\1&0&0}{\tau} + 20 \IMF{4&0&0\\3&0&0}{\tau} - 
  2 \IMF{4&0&0\\4&0&0}{\tau}\big]\,,\\
\nonumber M_{2}(x(\tau)) &=-\frac{1}{9 \pi ^3}
  \big[162 \IMF{4\\0}{\tau} - 180 \IMF{4\\1}{\tau} + 
   20 \IMF{4\\3}{\tau} - 2\IMF{4\\4}{\tau}  \\
\nonumber   &\qquad\qquad+ 
   243\IMF{4&0\\0&0}{\tau} - 270\IMF{4&0\\1&0}{\tau}+ 
   30 \IMF{4&0\\3&0}{\tau}- 3 \IMF{4&0\\4&0}{\tau}\big]\, ,\\
  \label{eq:solutionbanana}  M_{3}(x(\tau)) &=\frac{i}{18 \pi ^3}
   \big[81 \IMF{4\\0}{\tau}- 90 \IMF{4\\1}{\tau}+ 10 \IMF{4\\3}{\tau} - 
  \IMF{4\\4}{\tau}\big]\,.
\end{align}
We note again that the integrals $\IMF{n\\0}{\tau}$ are formally logarithmically
divergent for $\tau\to i\infty$. However, all divergences can be subtracted and
shuffled out in the standard way such that they are captured solely in terms of powers of $\IMF{0\\0}{\tau}=\tau$.
After regularisation, all iterated integrals of modular forms can be evaluated numerically with high precision, and we have checked that eq.~\eqref{eq:solutionbanana} numerically agrees with a direct numerical evaluation of the corresponding Feynman parameter representation.

We can see that, just like the result for the sunrise integral in
eq.~\eqref{eq:finalresult-sunrise}, our results in eq.~\eqref{eq:solutionbanana} have uniform weight one. 
Unlike the sunrise result, however, the
expressions for $M_i$ in eq.~\eqref{eq:solutionbanana} do not have uniform
length, i.e.~they are composed of iterated integrals with numbers of
integrations ranging from one to three. It is possible to perform a change of basis 
which casts the result as integrals which have both uniform length and
weight. In order to achieve this, we decompose the fundamental solution $\cW$
into a semi-simple times a unipotent matrix,

\begin{equation}
\label{eq:WSU}
  \cW=S\,U \,.
\end{equation}
An additional motivation to split the homogeneous solution into a semi-simple and a unipotent part comes from ref.~\cite{Broedel:2018qkq}, where it was argued that this splitting naturally leads to Feynman integrals of uniform weight.
An algorithmic way to construct this splitting in the present is described in appendix~\ref{app:semipotent}. Given the solution
matrix~$\cW$ and using eq.~\eqref{eq:psi3def}, we can find the unipotent matrix 
\begin{align}
\begin{split}
  \label{eqn:semisimpleunipotent}
  U&=
 \left(
\begin{array}{ccc}
 1 & -\frac{i \Psi_1(t)+\Psi_2(t))}{\Psi_1(t)} & -\frac{(\Psi_1(t)+\Psi_2(t)) (\Psi_1(t)+3 \Psi_2(t))}{\Psi_1(t)^2} \\
 0 & 1 & -\frac{2 i (2 \Psi_1(t)+3 \Psi_2(t))}{\Psi_1(t)} \\
 0 & 0 & 1 \\
\end{array}
\right)\\&=
  \left(
\begin{array}{ccc}
 1 & -i (\tau +1) & -(\tau +1) (3 \tau +1) \\
 0 & 1 & -2 i (3 \tau +2) \\
 0 & 0 & 1 \\
\end{array}
\right)\,.
\end{split}
\end{align}
Using this decomposition, we find that 
\begin{align}
\cW(x(\tau))\left(\begin{matrix} M_{1}(x(\tau)) \\ M_{2}(x(\tau))\\ M_{3}(x(\tau)) \end{matrix} \right)\,=\, S(\tau) \left(\begin{matrix} \tilde{M}_{1}(x(\tau)) \\ \tilde{M}_{2}(x(\tau))\\ \tilde{M}_{3}(x(\tau)) \end{matrix} \right)\ ,
\end{align}
with
\beq
 \left(\begin{matrix} \tilde{M}_{1}(x(\tau)) \\ \tilde{M}_{2}(x(\tau))\\ \tilde{M}_{3}(x(\tau)) \end{matrix} \right) = U(\tau)  \left(\begin{matrix} M_{1}(x(\tau)) \\ M_{2}(x(\tau))\\ M_{3}(x(\tau)) \end{matrix} \right) \ .
\eeq
The functions $\tilde{M}_i$ are of uniform weight two and of uniform length, given by
\begin{align}
\begin{split}
 \label{eq:MT}
    \tilde{M}_1(x(\tau)) &= -\frac{4\zeta_3}{\pi^2}-\frac{i}{3 \pi^3} 
  \big(81 \IMF{0&0&4 \\ 0 &0 &0}{\tau}-90 \IMF{0&0&4 \\ 0&0&1}{\tau }+10 \IMF{0&0&4 \\ 0&0&3}{\tau }-\IMF{0&0&4 \\ 0&0&4}{\tau }\big) \,,\\
  \tilde{M}_2(x(\tau)) &= \frac{1}{3 \pi^3} \,\big(81 \IMF{0&4 \\ 0&0}{\tau }-90 \IMF{0&4 \\ 0&1}{\tau }+10 \IMF{0&4 \\ 0&3}{\tau }-\IMF{0&4 \\ 0&4}{\tau }\big) \,,\\
\tilde{M}_3(x(\tau)) &= \frac{i}{18 \pi^3}  \big(81 \IMF{4 \\ 0}{\tau }-90 \IMF{4 \\ 1}{\tau }+10 \IMF{4 \\ 3}{\tau }-\IMF{4 \\ 4}{\tau } \big)
 \,.
 \end{split}
\end{align}
We can further see that the three different solutions
$\tilde{M}_i$ are related to each other by taking $\tau$-derivatives: 
\beq\bsp
 \tilde{M}_2(x(\tau)) &\,= i\partial_\tau  \tilde{M}_1(x(\tau))\,,\\
  \tilde{M}_3(x(\tau)) &\,= \frac{i}{6}\partial_\tau  \tilde{M}_2(x(\tau)) = -\frac{1}{6}\partial^2_\tau  \tilde{M}_1(x(\tau))\,.
\esp\eeq
In eq.~\eqref{eq:MT} we have expressed the banana integral
family in terms of iterated integrals of the same modular forms already
encountered in the sunrise case. It is worth exploring whether we can also
represent the solution in terms of another class of functions which play a
prominent role in the analytic calculation of the two-loop sunrise graph: elliptic multiple
polylogarithms (eMPLs). In the following we show how we can recast
eq.~\eqref{eq:MT} in terms of these functions.

\vspace{1mm}\noindent



\subsection{Solution in terms of elliptic polylogarithms}
\label{sec:BananaeMPLs}
In the previous section, we saw how the banana integral can be expressed in
terms of iterated integrals over the homogeneous solution of the sunrise
integral. The sunrise integral itself has been computed in many different forms
before. Here, we are particularly interested in the fact that the sunrise integral can also be represented in
terms of elliptic polylogarithms~\cite{Broedel:2017siw}.
It is therefore natural to ask the question whether it is possible to
express the banana integral in terms of elliptic polylogarithms as well.
To answer this question, let us recall the definition of the eMPLs as used in
ref.~\cite{Broedel:2017siw} (see also ref.~\cite{BrownLevin}),
\begin{align}
\sGt \left(\smg{n_1 & \dots
    & n_{k} \\ z_1 & \dots & z_{k} } ;z_{k+1}, \tau \right) = \int_0^{z_{k+1}}\!dw \, g^{(n_1)}(w-z_1;\tau) \,\sGt \left(\smg{n_2 & \dots
    & n_{k} \\ z_2 & \dots & z_{k} } ;w, \tau \right) \,.
\end{align}
Here the integration kernels $g^{(n)}(z;\tau)$ are related to expansion coefficients of the
Eisenstein-Kronecker series as defined in ref.~\cite{Broedel:2017siw}. The exact
form of these kernels is immaterial for the following arguments, though it is
important to note that for $z=\tfrac{r}{N}+\tfrac{s}{N}\tau$, with $r,s \in
\mathbb{Z}$ and $N \in \mathbb{N}$ the integration kernels $g$ can be expressed
as 
\begin{equation}
    g^{(n)}(\tfrac{r}{N}+\tfrac{s}{N}\tau,\tau) = \sum_{k=0}^{n}\frac{(2\pi
    i \tfrac{s}{N})^k}{k!}h^{(n-k)}_{N,r,s}(\tau)\,,
\end{equation}
where the functions $h^{(n-k)}_{N,r,s}$, $0\le r,s<N$, denote modular forms of weight $k>1$ for $\Gamma(N)$ (cf.~eq.~\eqref{eq:congruence_subgroups}) defined
as~\cite{Broedel:2018iwv}
\begin{equation}
	\label{eq:defEisenstein}
h^{(k)}_{N,r,s}(\tau) =-\!\!\!\!\!\!\!\sum_{\substack{(\alpha,\beta)\in
    \mathbb{Z}^2\\ (\alpha,\beta)\neq(0,0)}}\frac{e^{2\pi
    i(s\alpha-r\beta)/N}}{(\alpha+\beta\tau)^{2n}}\,.
\end{equation}
Not all these Eisenstein series are linearly independent. 
In ref.~\cite{Broedel:2018iwv} it was shown that the Eisenstein series of weight
$k\ge 2$ for $\Gamma(N)$ are spanned by the set $\{h^{(k)}_{N,r,s}\}_{0\le{}r,s<N}$.

From this observation it follows that whenever all the arguments of an eMPL are rational points, $z_i=\frac{r_i}{N}+\tau \frac{s_i}{N}$, then this function can be written as a linear combination of Eisenstein series for $\Gamma(N)$, defined as
\begin{equation}
\begin{split}
I\!\left(\begin{smallmatrix} n_1& N_1\\ r_1& s_1\end{smallmatrix}\big|\ldots\big|\begin{smallmatrix} n_k& N_k\\ r_k& s_k\end{smallmatrix};\tau\right) &\,\equiv I(h_{N_1,r_1,s_1}^{(n_1)},\ldots,h_{N_k,r_k,s_k}^{(n_k)};\tau)\\
&\,= \int_{i\infty}^{\tau}d\tau'\,h_{N_1,r_1,s_1}^{(n_1)}(\tau')\,I\left(\begin{smallmatrix} n_2& N_2\\ r_2& s_2\end{smallmatrix}\big|\ldots\big|\begin{smallmatrix} n_k& N_k\\ r_k& s_k\end{smallmatrix};\tau'\right)\,, \label{eq:itmod}
\end{split}
\end{equation}
with $h_{0,0,0}^{(0)}(\tau)\equiv 1$.  The converse statement,
however, is not always true: not every iterated integral of Eisenstein series
for $\Gamma(N)$ can be written in terms of eMPLs evaluated at rational points,
but this is only possible for specific combinations of iterated integrals of
Eisenstein series (cf., e.g., ref.~\cite{Pollack,Brown:mmv,Broedel:2015hia}).
The combination of iterated integrals of Eisenstein series that describes the
sunrise integrals satisfies this criterion. It is therefore natural to ask if
the same holds true for the banana graph. In ref.~\cite{Bloch:2014qca} it was
argued that the banana integral with unit powers of the propagators corresponds
to an elliptic trilogarithm. In the remainder of this section we make this
statement explicit and extend it to the other two master integrals for the
banana graph, and we show how the representation in terms of eMPLs can be
obtained. 

In broad strokes, the strategy we follow is very simple: we write down a very
general ansatz of eMPLs of length three with rational arguments with $N=6$. We
can express each of these in terms of iterated integrals of Eisenstein series
for $\Gamma(6)$ using the techniques described in ref.~\cite{Broedel:2018iwv},
and we match this expression to our results for the banana integrals from the
previous section. At this point, however, we need to make a technical
comment: while eMPLs naturally give rise to iterated integrals of
Eisenstein series for $\Gamma(6)$, the banana integrals in eq.~\eqref{eq:MT}
involve Eisenstein series for $\Gamma_1(6)$. Matching our ansatz of eMPLs to
eq.~\eqref{eq:MT} is therefore not completely straightforward. However, since
$\Gamma_1(6)$ is a subgroup of $\Gamma(6)$, we can express all Eisenstein
series for $\Gamma_1(6)$ in terms of those for $\Gamma(6)$. In particular, at
weight four (which is of relevance here, cf.~eq.~\eqref{eq:MT}), there are
four Eisenstein series for $\Gamma_1(6)$, which can be written as linear
combinations of the basis of Eisenstein series for $\Gamma(6)$ as follows,
\begin{equation}
\begin{array}{l}
b_1(\tau)=h^{(4)}_{6,0,1}(\tau)+h^{(4)}_{6,1,1}(\tau)+h^{(4)}_{6,2,1}(\tau)+h^{(4)}_{6,3,1}(\tau)+h^{(4)}_{6,4,1}(\tau)+h^{(4)}_{6,5,1}(\tau)\, ,\\
b_2(\tau)=h^{(4)}_{6,1,2}(\tau)+h^{(4)}_{6,3,2}(\tau)+h^{(4)}_{6,5,2}(\tau)\, , \\
b_3(\tau)=h^{(4)}_{6,1,0}(\tau)\, ,\\
b_4(\tau)=h^{(4)}_{6,1,3}(\tau)+h^{(4)}_{6,4,3}(\tau)\,.
\end{array}
\end{equation}
Alternatively, the elements $b_i(\tau)$ can be expressed in terms of the
funtions $f_{n,p}(\tau)$ defined in eq.~\eqref{eq:Gamma_1(6)_basis},
\begin{equation}
 \label{eq:bfnp}
\begin{array}{l}
	b_1(\tau)= -\frac{91}{2880} f_{4,0}(\tau) +\frac{3}{80}  f_{4,1}(\tau) -\frac{7}{864} f_{4,2}(\tau)+\frac{1}{2160}f_{4,3}(\tau) +\frac{7}{77760}  f_{4,4}(\tau)\,, \\[5pt]
	b_2(\tau)=  \frac{13}{720}  f_{4,0}(\tau) -\frac{1}{180} f_{4,1}(\tau) +\frac{1}{216} f_{4,2}(\tau)-\frac{1}{180} f_{4,3}(\tau) -\frac{1}{19440}  f_{4,4}(\tau) \,, \\[5pt]
	b_3(\tau)= -\frac{1}{80}    f_{4,0}(\tau) -\frac{3}{20}  f_{4,1}(\tau) +\frac{1}{72}  f_{4,2}(\tau)-\frac{1}{540} f_{4,3}(\tau) +\frac{13}{19440} f_{4,4}(\tau) \,, \\[5pt]
	b_4(\tau)=  \frac{7}{320}   f_{4,0}(\tau) +\frac{1}{80}  f_{4,1}(\tau) -\frac{7}{288} f_{4,2}(\tau)+\frac{1}{80}  f_{4,3}(\tau) -\frac{91}{77760} f_{4,4}(\tau)\,.
\end{array}
\end{equation}
Using these relations, we can express every iterated integral in
eq.~\eqref{eq:MT} in terms of the iterated integrals defined in
eq.~\eqref{eq:itmod}. We find
\begin{align}
 \nonumber \tilde{M}_1(x(\tau))&=\frac{36i}{\pi^3}\Big(21\,I(\smz|\smz|\sm{0}{1};\tau) - \,I(\smz| \smz| \sm{1}{0};\tau) + 21\,I(\smz| \smz| \sm{1}{1};\tau)\\
\nonumber	   &\qquad\quad+ 3\,I(\smz| \smz| \sm{1}{2};\tau) - 7\,I(\smz| \smz| \sm{1}{3};\tau) + 21\,I(\smz| \smz| \sm{2}{1};\tau) \\
\nonumber    &\qquad\quad+ 21\,I(\smz| \smz| \sm{3}{1};\tau) + 3\,I(\smz| \smz| \sm{3}{2};\tau) + 21\,I(\smz| \smz| \sm{4}{1};\tau) \\
 \nonumber   &\qquad\quad- 7\,I(\smz| \smz| \sm{4}{3};\tau) + 21\,I(\smz| \smz| \sm{5}{1};\tau) + 3\,I(\smz| \smz| \sm{5}{2};\tau)\Big)\\
 \nonumber   &\qquad\quad - \frac{4 \zeta_3}{\pi^2},\\
 \nonumber \tilde{M}_2(x(\tau))&=-\frac{36}{\pi^3}\Big(21\,I(\smz| \sm{0}{1};\tau) - \,I(\smz| \sm{1}{0};\tau) + 21\,I(\smz| \sm{1}{1};\tau) \\
\nonumber	 &\qquad\quad + 3\,I(\smz| \sm{1}{2};\tau) - 7\,I(\smz| \sm{1}{3};\tau) + 21\,I(\smz| \sm{2}{1};\tau) \\
\nonumber  &\qquad\quad + 21\,I(\smz| \sm{3}{1};\tau) + 3\,I(\smz| \sm{3}{2};\tau) + 21\,I(\smz| \sm{4}{1};\tau) \\
\nonumber  &\qquad\quad- 7\,I(\smz| \sm{4}{3};\tau) + 21\,I(\smz| \sm{5}{1};\tau) + 3\,I(\smz| \sm{5}{2};\tau)\Big),\\
 \nonumber \tilde{M}_3(x(\tau))&= -\frac{6i}{\pi^3}\Big(21\,I(\sm{0}{1};\tau) - \,I(\sm{1}{0};\tau) +21\,I(\sm{1}{1};\tau) +3\,I(\sm{1}{2};\tau) \\
\nonumber	 &\quad\qquad - 7\,I(\sm{1}{3};\tau) +21\,I(\sm{2}{1};\tau) +21\,I(\sm{3}{1};\tau) +
  3\,I(\sm{3}{2};\tau) \\
  \label{eq:solution}&\quad\qquad+ 21\,I(\sm{4}{1};\tau) -7 \,I(\sm{4}{3};\tau) +21 \,I(\sm{5}{1};\tau) +3 \,I(\sm{5}{2};\tau)\Big)\,.
\end{align}
This result allows us to make the connection to eMPLs.  In order to find a
representation of the $\tilde{M}_i$ in terms of eMPLs, we write a suitable
ansatz for them in terms of eMPLs, rewrite these eMPLs in terms of iterated
integrals of Eisenstein series and then fix the coefficients in the ansatz with
the results given in eq.~\eqref{eq:solution}. The ansatz we have chosen mirrors
the observation that the $\tilde{M}_i$ are linear combinations of iterated
integrals of modular forms of uniform weight and length, that all iterated
integrals have leading zero entries and that only the last entry is a
nontrivial modular form. Consequently, we chose elliptic multiple
polylogarithms of the form
\begin{align}
\sGt \Big( \hspace{-0.2cm}\underbrace{\begin{smallmatrix} 0 & \dots & 0 \\ 0 & \dots & 0 \end{smallmatrix}}_{k-\text{times}}  \hspace{-0.2cm} \smg{3-k\\z_1};z,\tau \Big) 
\end{align}
evaluated at rational points such that all solutions are related by adding or
removing leading zero entries. More precisely, our ansatz consists of all
antisymmetric combinations 
\begin{align}
\Gamma_{\text{AS}}^{k}\left( \smg{r_1 &  r_2\\ s_1 & s_2} \right) = \sGt \Big( \hspace{-0.5cm}\underbrace{\begin{smallmatrix} 0 & \cdots & 0 \\ 0 & \cdots & 0 \end{smallmatrix}}_{(3-k)-\text{times}}  \hspace{-0.5cm} \smg{k\\z_1};z_2,\tau \Big) - \sGt \Big( \hspace{-0.5cm}\underbrace{\begin{smallmatrix} 0 & \cdots & 0 \\ 0 & \cdots & 0 \end{smallmatrix}}_{(3-k)-\text{times}}  \hspace{-0.5cm} \smg{k\\-z_1};z_2,\tau \Big) \,,
\end{align}
with $z_i = \frac{r_i}{6} + \frac{s_i}{6} \tau$ for $r_i,s_i \in \lbrace 0,\dots,5 \rbrace$ and for $1 \leq k \leq 3$. 

We find that the solutions $\tilde{M}_i$ can indeed be expressed in terms of
elliptic multiple polylogarithms and a possible representation is given by (for
$k=1,2,3$)
\begin{align}
    \tilde{M}_k = C_k\tilde{m}_k\,,
\end{align}
where the prefactors are given by

\begin{align}
C_1 \,=\, -12 \,,\quad C_2 \,=\, -\frac{6}{ \pi}\,,\quad C_3 \,=\, -\frac{1}{\pi^2} \,,
\end{align}
and the functions $m_k$ are defined as
\begin{align}\label{eq:solutionEMPL}
\tilde{m}&_k=-\frac{13319}{96}\sGtAS^{k}\left(\smg{0 & 3 \\ 1 & 0} \right)+\frac{2679}{160}\sGtAS^{k}\left(\smg{0 & 5 \\ 1 & 4} \right)-\frac{77}{10}\sGtAS^{k}\left(\smg{0 & 3 \\ 2 & 0} \right)\\ \nonumber
&-\frac{2911}{15}\sGtAS^{k}\left(\smg{0 & 0 \\ 3 & 2} \right)+\frac{20261}{1440}\sGtAS^{k}\left(\smg{0 & 1 \\ 3 & 4} \right)+\frac{577}{60}\sGtAS^{k}\left(\smg{0 & 2 \\ 3 & 5} \right)-\frac{22841}{120}\sGtAS^{k}\left(\smg{0 & 4 \\ 3 & 1} \right)\\ \nonumber
&+\frac{1639}{180}\sGtAS^{k}\left(\smg{0 & 4 \\ 3 & 5} \right)-\frac{755827}{7200}\sGtAS^{k}\left(\smg{0 & 5 \\ 3 & 0} \right)-\frac{1371547}{2160}\sGtAS^{k}\left(\smg{0 & 5 \\ 3 & 2} \right)+\frac{969431}{720}\sGtAS^{k}\left(\smg{0 & 5 \\ 3 & 3} \right)\\ \nonumber
&-\frac{1011209}{2160}\sGtAS^{k}\left(\smg{0 & 5 \\ 3 & 5} \right)+\frac{77}{20}\sGtAS^{k}\left(\smg{0 & 3 \\ 4 & 0} \right)-\frac{70291}{480}\sGtAS^{k}\left(\smg{0 & 3 \\ 5 & 0} \right)+\frac{2679}{160}\sGtAS^{k}\left(\smg{0 & 5 \\ 5 & 4} \right)\\ \nonumber
&-\frac{10409}{90}\sGtAS^{k}\left(\smg{1 & 0 \\ 0 & 3} \right)+\frac{2197}{300}\sGtAS^{k}\left(\smg{1 & 5 \\ 0 & 5} \right)-\frac{893}{120}\sGtAS^{k}\left(\smg{1 & 0 \\ 1 & 3} \right)+\frac{665}{6}\sGtAS^{k}\left(\smg{1 & 3 \\ 1 & 0} \right)\\ \nonumber
&-\frac{57739}{288}\sGtAS^{k}\left(\smg{1 & 0 \\ 2 & 1} \right)+\frac{36031}{1440}\sGtAS^{k}\left(\smg{1 & 0 \\ 2 & 5} \right)-\frac{140}{3}\sGtAS^{k}\left(\smg{1 & 3 \\ 2 & 0} \right)+\frac{14}{5}\sGtAS^{k}\left(\smg{1 & 3 \\ 2 & 3} \right)\\ \nonumber
&+\frac{22867}{360}\sGtAS^{k}\left(\smg{1 & 3 \\ 3 & 0} \right)-\frac{2069}{40}\sGtAS^{k}\left(\smg{1 & 4 \\ 3 & 3} \right)-\frac{1427}{40}\sGtAS^{k}\left(\smg{1 & 0 \\ 4 & 3} \right)+\frac{847}{40}\sGtAS^{k}\left(\smg{1 & 0 \\ 4 & 4} \right)\\ \nonumber
&+\frac{7343}{60}\sGtAS^{k}\left(\smg{1 & 3 \\ 4 & 0} \right)-\frac{1579}{120}\sGtAS^{k}\left(\smg{1 & 0 \\ 5 & 3} \right)-\frac{55}{8}\sGtAS^{k}\left(\smg{1 & 0 \\ 5 & 4} \right)+\frac{6207}{40}\sGtAS^{k}\left(\smg{2 & 0 \\ 0 & 3} \right)\\ \nonumber
&-\frac{386267}{720}\sGtAS^{k}\left(\smg{2 & 0 \\ 0 & 4} \right)-\frac{2197}{40}\sGtAS^{k}\left(\smg{2 & 1 \\ 0 & 0} \right)-\frac{386267}{360}\sGtAS^{k}\left(\smg{2 & 4 \\ 0 & 2} \right)+\frac{386267}{360}\sGtAS^{k}\left(\smg{2 & 4 \\ 0 & 4} \right)\\ \nonumber
&-\frac{72913}{360}\sGtAS^{k}\left(\smg{2 & 3 \\ 3 & 0} \right)+\frac{1481}{20}\sGtAS^{k}\left(\smg{2 & 3 \\ 3 & 3} \right)+\frac{665}{12}\sGtAS^{k}\left(\smg{2 & 3 \\ 4 & 0} \right)+\frac{893}{60}\sGtAS^{k}\left(\smg{2 & 0 \\ 5 & 3} \right)\\ \nonumber
&-\frac{1367}{30}\sGtAS^{k}\left(\smg{3 & 0 \\ 3 & 5} \right)-\frac{188113}{10800}\sGtAS^{k}\left(\smg{3 & 2 \\ 3 & 0} \right)+\frac{105}{2}\sGtAS^{k}\left(\smg{3 & 3 \\ 3 & 1} \right)+\frac{263}{3}\sGtAS^{k}\left(\smg{3 & 3 \\ 3 & 2} \right)\\ \nonumber
&+\frac{1582769}{10800}\sGtAS^{k}\left(\smg{3 & 4 \\ 3 & 0} \right)-\frac{1555}{8}\sGtAS^{k}\left(\smg{3 & 5 \\ 3 & 0} \right)+\frac{77}{10}\sGtAS^{k}\left(\smg{3 & 3 \\ 4 & 0} \right)+\frac{203}{30}\sGtAS^{k}\left(\smg{3 & 0 \\ 5 & 3} \right)\\ \nonumber
&-\frac{21}{2}\sGtAS^{k}\left(\smg{3 & 3 \\ 5 & 3} \right)+\frac{14}{5}\sGtAS^{k}\left(\smg{3 & 3 \\ 5 & 4} \right)+\frac{8141}{120}\sGtAS^{k}\left(\smg{4 & 0 \\ 1 & 3} \right)+\frac{1271}{24}\sGtAS^{k}\left(\smg{4 & 3 \\ 1 & 0} \right)\\ \nonumber
&-\frac{1271}{24}\sGtAS^{k}\left(\smg{4 & 3 \\ 1 & 3} \right)-\frac{386267}{720}\sGtAS^{k}\left(\smg{4 & 0 \\ 2 & 2} \right)+\frac{386267}{720}\sGtAS^{k}\left(\smg{4 & 2 \\ 2 & 0} \right)+\frac{386267}{720}\sGtAS^{k}\left(\smg{4 & 2 \\ 2 & 2} \right)\\ \nonumber
&-\frac{665}{6}\sGtAS^{k}\left(\smg{4 & 3 \\ 2 & 0} \right)-\frac{386267}{720}\sGtAS^{k}\left(\smg{4 & 4 \\ 2 & 0} \right)-\frac{31277}{360}\sGtAS^{k}\left(\smg{4 & 3 \\ 3 & 0} \right)-\frac{147}{5}\sGtAS^{k}\left(\smg{4 & 3 \\ 3 & 3} \right)\\ \nonumber
&-\frac{386267}{360}\sGtAS^{k}\left(\smg{4 & 2 \\ 4 & 0}
    \right)+\frac{386267}{720}\sGtAS^{k}\left(\smg{4 & 4 \\ 4 & 0}
    \right)-\frac{253}{180}\sGtAS^{k}\left(\smg{4 & 0 \\ 5 & 3}
    \right)-\frac{1111}{10}\sGtAS^{k}\left(\smg{4 & 3 \\ 5 & 0}
    \right)\\\nonumber
\phantom{C_k \tilde{M}}&-\frac{41}{12}\sGtAS^{k}\left(\smg{4 & 3 \\ 5 & 2} \right)+\frac{221}{60}\sGtAS^{k}\left(\smg{4 & 3 \\ 5 & 3} \right)+\frac{48}{5}\sGtAS^{k}\left(\smg{5 & 0 \\ 1 & 1} \right)+\frac{519}{40}\sGtAS^{k}\left(\smg{5 & 0 \\ 1 & 3} \right)\\ \nonumber
&-\frac{77}{2}\sGtAS^{k}\left(\smg{5 & 3 \\ 1 & 3} \right)+\frac{70}{}\sGtAS^{k}\left(\smg{5 & 4 \\ 1 & 0} \right)+\frac{168}{5}\sGtAS^{k}\left(\smg{5 & 5 \\ 1 & 0} \right)-\frac{1045553}{720}\sGtAS^{k}\left(\smg{5 & 0 \\ 2 & 2} \right) \\ \nonumber
&+\frac{231407}{270}\sGtAS^{k}\left(\smg{5 & 0 \\ 2 & 3}
    \right)+\frac{1127}{60}\sGtAS^{k}\left(\smg{5 & 3 \\ 2 & 0}
    \right)+\frac{518}{5}\sGtAS^{k}\left(\smg{5 & 3 \\ 2 & 3}
    \right)+\frac{2069}{40}\sGtAS^{k}\left(\smg{5 & 0 \\ 3 & 3}
    \right)\\\nonumber
    &-\frac{9379}{360}\sGtAS^{k}\left(\smg{5 & 3 \\ 3 & 0} \right)-\frac{126}{5}\sGtAS^{k}\left(\smg{5 & 4 \\ 3 & 0} \right)+\frac{126}{5}\sGtAS^{k}\left(\smg{5 & 5 \\ 3 & 0} \right)-\frac{21637}{160}\sGtAS^{k}\left(\smg{5 & 0 \\ 4 & 1} \right)\\ \nonumber
&-\frac{1579}{160}\sGtAS^{k}\left(\smg{5 & 0 \\ 4 & 5} \right)+\frac{518}{5}\sGtAS^{k}\left(\smg{5 & 1 \\ 4 & 1} \right)+\frac{223}{10}\sGtAS^{k}\left(\smg{5 & 3 \\ 4 & 0} \right)+\frac{24}{5}\sGtAS^{k}\left(\smg{5 & 4 \\ 4 & 0} \right)\\ \nonumber
&+\frac{14}{5}\sGtAS^{k}\left(\smg{5 & 4 \\ 4 & 5} \right)-\frac{208783}{144}\sGtAS^{k}\left(\smg{5 & 0 \\ 5 & 1} \right)+\frac{1078601}{2160}\sGtAS^{k}\left(\smg{5 & 0 \\ 5 & 3} \right)-\frac{141}{10}\sGtAS^{k}\left(\smg{5 & 0 \\ 5 & 5} \right)\\ \nonumber
&-\frac{37841}{2700}\sGtAS^{k}\left(\smg{5 & 2 \\ 5 & 0} \right)+\frac{321817}{2700}\sGtAS^{k}\left(\smg{5 & 3 \\ 5 & 0} \right)-\frac{136121}{2700}\sGtAS^{k}\left(\smg{5 & 4 \\ 5 & 0} \right)+\frac{12277}{300}\sGtAS^{k}\left(\smg{5 & 5 \\ 5 & 0} \right) \,.
\end{align}
We note here that the boundary constant proportional to $\zeta_3$, appearing in eq.~\eqref{eq:solution} does not appear explicitly in this
representation, as for $k = 1$ the term proportional to $\zeta_3$ is
contained in the combination of eMPLs.

\vspace{1mm}\noindent



\section{Conclusion and Outlook}
\label{sec:conclusions}

In this paper we have presented for the first time fully analytic results for
all master integrals of the equal-mass three-loop banana graph. Our results are
characterised by remarkable simplicity, and they only involve the same class of
functions that shows up also in the two-loop equal-mass sunrise graph, namely
iterated integrals for modular forms for $\Gamma_1(6)$ and elliptic
polylogarithms evaluated at rational points. 

Our paper is also the first time that a family of Feynman integral whose
associated Picard-Fuchs operator is irreducible of order three has been
evaluated analytically in terms of a well-established class of transcendental
functions.  This result may have important implications for tackling
phenomenologically relevant three-loop processes involving massive virtual
particles.  In particular, the banana graph is the simplest subtopology that
appears in the computation of the three-loop corrections to Higgs production
via gluon fusion where the dependence on the top-quark mass is kept. While
these corrections are known numerically~\cite{Davies:2019nhm}, no analytic
solution is known. 
Correspondingly, the full analytic result will necessarily involve integrals
over the banana graphs. Our results in terms of iterated integrals of modular
forms are well suited to perform these integrals.  Most likely, however, also
higher orders in the $\eps$-expansion of the banana graph would be required and
we expect that the techniques presented in this paper can be extended to this
case as well. This is left for future work.

\vspace{1mm}\noindent

\appendix


\section{Boundary condition for the banana graph}
\label{app:initial_condition}
In this section we discuss how to obtain the leading asymptotic expansion of the master integrals for the banana integrals in eq.~\eqref{eq:mba}.
Asymptotic expansions for Feynman integrals are a well studied topic in the context of the method of expansion-by-regions~\cite{regions1,regions2}. Here we will employ a particular method that relies on Mellin-Barnes integral transformations to obtain the asymptotic expansion of the banana Feynman integrals around the point $x=0$.

We start by Feynman parametrising the integral. The Symanzik polynomials relevant for the three master integrals in eq.~\eqref{eq:fints} are
\begin{equation}
    \label{eq:symanzik}
    \begin{split}
	    \mathcal{U} &= x_1 x_2 x_3+x_1 x_4 x_3+x_2 x_4 x_3+x_1 x_2 x_4\quad\textrm{and}\\
	    \mathcal{F} &= s\,x_1x_2x_3x_4 + m^2 (x_1+x_2+x_3+x_4)\,\mathcal{U}\,,
    \end{split}
\end{equation}
so that we can write the Feynman parametric representation of the first master integral as
\begin{align} \begin{split}
    \cI_1\,=\,& (1+2\eps)(1+3\eps)I_{1,1,1,1,0,0,0,0,0}\\
    \,=\,&\frac{(1+2\eps)(1+3\eps)}{\Gamma(1+\eps)^3}\int\,dx_1dx_2dx_3dx_4\delta(1-x_4)\,\mathcal{U}^{4\eps}\,\mathcal{F}^{-1-3\eps}\,,
\end{split} \end{align}
where we have chosen the argument of the $\delta$ function in a way that is
advantageous for the rest of the calculation.

We can use a useful trick to simplify the integral drastically, at the cost of
introducing an additional integration: we introduce a Mellin-Barnes parameter
by using the identity,
\begin{equation}
    \label{eq:mbtrick}
    \frac{1}{(A+B)^{\lambda}} = \int_{\mathcal{C}-i\infty}^{\mathcal{C}+i\infty}\frac{d\xi}{2\pi i} A^{\xi}B^{-\xi-\lambda}\frac{\Gamma(-\xi)\Gamma(\xi+\lambda)}{\Gamma(\lambda)}\,,
\end{equation}
where the contour of integration runs parallel to the imaginary axis and
intersects the real axis at a point $\mathcal{C}$ that is chosen such that the
contour separates the left poles of the integrand (due to
$\Gamma(\xi+\lambda)$) from the right poles (due to $\Gamma(-\xi)$).
We can use this identity to separate the two terms in the $\mathcal{F}$
polynomial and write the integral as
\begin{equation}
    \begin{split}
      \cI_1=\int&\frac{d\xi_1}{2\pi i}\frac{\Gamma(-\xi_1)\Gamma(1+3\eps+\xi_1)}{\Gamma(1+\eps)^3}\left(-x/4\right)^{-\xi_1}\\
    \times\int&\,dx_1dx_2dx_3dx_4\delta(1-x_4)\left(x_1x_2x_3x_4\right)^{\xi_1}\left(x_1+x_2+x_3+x_4\right)^{-1-3\eps-\xi_1}\mathcal{U}^{-1+\eps-\xi_1}\,.
\end{split}
\end{equation}
This transformation renders the integral effectively massless and we can proceed to integrate out the Feynman parameters $x_i$ one at a time.
In doing so we encounter two integrals of the form
\begin{equation}
    \int_0^{\infty}dx x^{\alpha}(A+B x)^{\beta}(C+D x)^{\gamma}\,.
\end{equation}
Ordinarily, such an integral can be evaluated in terms of hypergeometric functions. However, in this case it is advantageous to instead apply the Mellin-Barns trick from eq.~\eqref{eq:mbtrick} once more, in order to split one of the two linear terms into monomial factors, which will allow us to perform the integral in terms of $\Gamma$ functions as
\begin{equation}
    \int_0^{\infty}dx x^{\alpha}(A+B x)^{\beta} = A^{1+\alpha+\beta}B^{-1-\alpha}\frac{\Gamma(1+\alpha)\Gamma(-1-\alpha-\beta)}{\Gamma(-\beta)}\,.
\end{equation}
After integrating out the Feynman parameters in this fashion, we find the following Mellin-Barnes representation of the integral,
\begin{equation}
    \begin{split}
        \label{eq:lolrep}
        \mathcal{I}_1 =(1+2\eps)(1+3\eps)\int&\,\frac{d\xi_1d\xi_2d\xi_3}{(2\pi i)^3}\left(-x/4\right)^{-\xi_1}\Gamma(-\xi_1)\Gamma(-\xi_2)\Gamma(-\xi_3)\\
    \times\Gamma(\eps-\xi_2)&\Gamma(\eps-\xi_3)\frac{\Gamma(1+\xi_{123})^2\Gamma(1-\eps+\xi_{123})\Gamma(1+\eps+\xi_{123})}{\Gamma(1+\eps)^3\Gamma(2+2\xi_{123})\Gamma(1-\eps+\xi_1)}\,,
\end{split}
\end{equation}
where we have defined the abbreviation $\xi_{123} = \xi_1+\xi_2+\xi_3$. In the
above integral, the contour of integration is defined implicitly through the
requirement that it separates the left and right poles of the $\Gamma$
functions. An explicit representation of the contour can be obtained in an
algorithmic fashion as implemented for example in the \texttt{Mathematica}
packages \texttt{MB}~\cite{Czakon:2005rk} and
\texttt{MBresolve}~\cite{Smirnov:2009up}. The explicit form of the contour is
useful when using Cauchy's theorem to perform the remaining integrations, by
closing the contour and summing residues. However, we will see that this is
actually not necessary in this case.

First of all, so far we have not performed any asymptotic expansion and
eq.~\eqref{eq:lolrep} is a Mellin-Barnes representation of the entire integral,
but we only care about the integral in the limit $x\to0$.

The Mellin-Barnes representation allows us to take the asymptotic limit in a
straightforward fashion: inspecting the integrand of the Mellin-Barnes
representation we see that for generic values of the $\xi_i$ the integrand
vanishes when we take $x\to0$. To obtain the Laurent expansion around vanishing
$x$, we therefore need to take the leading residues, starting from $\xi_1=1$.
The surviving residues can be
determined algorithmically for example using the package
\texttt{MBasymptotics}~\cite{mbasymptotics}. Solving the constraints for our integral in
eq.~\eqref{eq:lolrep} we find that the only terms contributing in the
limit $x\to0$ are codimension three residues so that no integrations remain.
We have,
\begin{align}
  \nonumber      \frac{\lim_{x\to0}\cI_1}{(1+2\eps)(1+3\eps)}&=-\frac{x}{\eps^3}-\frac{12\,\Gamma(1-\eps)^2}{\eps^3\Gamma(1-2\eps)}(-x/4)^{1+\eps}+\frac{12\,\Gamma(1-\eps)^3\Gamma(1+2\eps)}{\eps^3\Gamma(1-3\eps)\Gamma(1+\eps)^2}(-x/4)^{1+2\eps}\\
 \nonumber   &\quad-\frac{4\,\Gamma(1-\eps)^4\Gamma(1+3\eps)}{\eps^3\Gamma(1-4\eps)\Gamma(1+\eps)^3}(-x/4)^{1+3\eps}\\
 \nonumber   &= x\big[\log(-x/4)^3-4\zeta_3\big]\\
  \nonumber      &\quad+\eps{}x\big[-6\zeta_4-20\zeta_3-30\zeta_3\log(-x/4)-3\zeta_2\log(-x/4)^2\\
        &\quad\quad\quad\phantom{\big[}\,\,+5\log(-x/4)^3+\tfrac{3}{2}\log(-x/4)^4\big]+\mathcal{O}(\eps^2)\,.
\end{align}
The other two master integrals can can be computed completely analogously, the
only difference are shifted exponents of the Symanzik polynomials, and we can
obtain the asymptotic limit for the second master integral as 
\begin{align}
\nonumber    \frac{\lim_{x\to0}\cI_2}{(1+2\eps)}&=\frac{3}{4\eps^2}x+\frac{6\Gamma(1-\eps)^2}{\eps^2\Gamma(1-2\eps)}(-x/4)^{1+\eps}-\frac{3\Gamma(1-\eps)^3\Gamma(1+2\eps)}{\eps^2\Gamma(1-3\eps)\Gamma(1+\eps)^2}(-x/4)^{1+2\eps}\\
 \nonumber   &=\big[\tfrac{3}{4}x\log(-x/4)^2\big]+\eps{}x\big[\tfrac{9}{2}\zeta_3-\tfrac{3}{2}\zeta_2\log(-x/4)+\tfrac{3}{2}x\log(-x/4)^2\\
    &\quad\quad+\tfrac{3}{4}\log(-x/4)^3\big]+\mathcal{O}(\eps^2)\,,
\end{align}
and similarly for the third master integral,
\begin{equation}
    \begin{split}
        \lim_{x\to0}\cI_3&=-\frac{1}{2\eps}x-\frac{2\Gamma(1-\eps)^2}{\eps\Gamma(1-2\eps)}(-x/4)^{1+\eps}\\
        &=\big[\tfrac{1}{2}x\log(-x/4)\big]+\eps
        x \big[-\tfrac{1}{2}\zeta_2+\tfrac{1}{4}\log(-x/4)^2\big]+\mathcal{O}(\eps^2)\,.
    \end{split}
\end{equation}


\section{Decomposing a matrix into a semi-simple and a unipotent part}
\label{app:semipotent}

In this appendix we show how to decompose an invertible matrix $\Omega$ (with certain additional conditions, see below)
into a product of a lower and and upper-triangular matrix. From this we can infer the decomposition of the period matrix of the banana graph into a semi-simple and a unipotent matrix, see eq.~\eqref{eq:WSU}.
 A unipotent matrix is a matrix whose
difference to the unit matrix is nilpotent. Good examples are upper triangular
matrices with only ones on the diagonal. A semi-simple matrix, on the other hand, is a matrix
which is similar to a direct sum of simple matrices. Over an algebraically
closed field (e.g., the complex numbers), semi-simple matrices are just the diagonalisable matrices. 

Let us define the matrix
\begin{align}
  \label{eq:periodmatrix}
  \Omega=\big(\Omega_{ij}\big)_{1\leq i,j \leq n}\,,
\end{align}
which we assume to be invertible. In the following we are going to show that
the matrix can be decomposed into the upper-triangular matrix $U$ and a lower-triangular
matrix $S$ such that 
\begin{equation}
  \label{eqn:}
  \Omega=S\,U\,.
\end{equation}
In order to do so, consider the {principle minors} of $\Omega$,
\begin{align}
  M_k=\det \big(\Omega_{ij}\big)_{1\leq i,j \leq k}=\det \Omega^{(k)}, \qquad M_0\equiv 1. 
\end{align}
Furthermore, let us define auxiliary sets of matrices:
\begin{align}
  O_{k\ell}=\det
\begin{pmatrix}
  \Omega_{11} & \cdots & \Omega_{1 k-1} & \Omega_{1\ell} \\
  \vdots      &        & \vdots         & \vdots   \\
  \Omega_{k1} & \cdots & \Omega_{k k-1} & \Omega_{k\ell} 
\end{pmatrix},\quad
  C_{k\ell}=\det
\begin{pmatrix}
  \Omega_{11}     & \cdots & \Omega_{1 k}    \\
  \vdots          &        & \vdots          \\
  \Omega_{k-11}   & \cdots & \Omega_{k-1 k}  \\
  \Omega_{\ell 1} & \cdots & \Omega_{\ell k} 
\end{pmatrix}.
\end{align}
Using those auxiliary objects, define the matrices:
\begin{equation}
  \label{eq:defUS}
  S_{ij}=\frac{C_{ji}}{M_{j-1}} \text{ and } U_{ij}=\frac{O_{ij}}{M_i}\,.
\end{equation}
Note that for the previous equation to make sense, we need to require that all principle minors of $\Omega$ be non-zero.

Writing out the product of $S$ and $U$, one finds 
\begin{align}
  \sum\limits_{k=1}^n S_{ik} U_{kj}&=\sum\limits_{k=1}^nC_{ki}O_{kj}\notag\\
				   &=\sum\limits_{k=1}^n\frac{1}{M_k M_{k-1}}\det
\begin{pmatrix}
  \Omega^{(k-1)}  & \Omega_{\ast k}    \\
  \Omega_{i\ast}  & \Omega_{ik} 
\end{pmatrix}\det
\begin{pmatrix}
  \Omega^{(k-1)}  & \Omega_{\ast j}    \\
  \Omega_{k\ast}  & \Omega_{kj} 
\end{pmatrix}\notag\\
&
=\sum\limits_{k=1}^n\frac{1}{\det \Omega^{(k)} \det\Omega^{(k-1)}}\det
\begin{pmatrix}
  \Omega^{(k-1)}  & \Omega_{\ast k}    \\
  \Omega_{i\ast}  & \Omega_{ik} 
\end{pmatrix}\det
\begin{pmatrix}
  \Omega^{(k-1)}  & \Omega_{\ast j}    \\
  \Omega_{k\ast}  & \Omega_{kj} 
\end{pmatrix}\notag\\
&=\Omega_{ij}\,. 
\end{align}
The manipulations in the above equation for generic matrices are algebraically rather involved. 
We have therefore limited ourselves to  testing explicitly the correctness of this formula 
for matrices up to
$n=10$.

Next, we note that the matrix $U$ has the following shape:
\begin{align}
\begin{pmatrix}
  1      &  & * \\
         & \ddots  &  \\
  0      &  &  1 
\end{pmatrix}\,.
\end{align}
This immediately implies that $U$ is unipotent.
Indeed, considering $i>j$, one finds
\begin{equation}
  U_{ij}=O_{ij}=\frac{1}{M_i}\det
\begin{pmatrix}
  \Omega_{11} & \cdots & \Omega_{1 i-1} & \Omega_{1j} \\
  \vdots      &        & \vdots         & \vdots   \\
  \Omega_{i1} & \cdots & \Omega_{i i-1} & \Omega_{ij} 
\end{pmatrix}=0\,.
\end{equation}
because there are two identical columns. For the diagonal elements one finds
\begin{equation}
  U_{ii}=O_{ii}=\frac{1}{M_i}M_i=1\,. 
\end{equation}
One can show along the same lines that all
elements $S_{ij}$ for $i<j$ vanish, and so $S$ is lower-triangular. 

The previous considerations do not yet allow us to conclude that $S$ is semi-simple, because not every lower-triangular matrix is diagonalisable. We can, however, easily check that the matrix $S$ obtained in this way is diagonalisable on a case by case basis. 
Indeed, a sufficient criterion for a triangular matrix to be diagonalisable is that all its diagonal elements are distinct (because in that case the matrix has a maximal number of distinct eigenvalues). 
In particular, we can then easily check that this construction leads to a semi-simple matrix $S$ in the case of the banana graph where $\Omega=\cW(x)$, with $\cW$ is defined in eq.~\eqref{eq:homsol}. Indeed, we immediately see that in that case $S$ has three distinct eigenvalues for generic values of $x$. Therefore, $S$ is diagonalisable for generic $x$, and hence semi-simple. We have thus obtained the desired decomposition into a semi-simple and a unipotent matrix.


\bibliographystyle{JHEP}
\bibliography{ellip}

\providecommand{\href}[2]{#2}\begingroup\raggedright\begin{thebibliography}{10}

\bibitem{Kummer}
E.~E. Kummer, {\it {\"{U}ber die Transcendenten, welche aus wiederholten
  Integrationen rationaler Formeln entstehen}},  {\em J. reine ang. Mathematik}
  {\bf 21} (1840) 74--90; 193--225; 328--371.

\bibitem{Nielsen}
N.~Nielsen, {\it {Der Eulersche Dilogarithmus und seine Verallgemeinerungen}},
  {\em Nova Acta Leopoldina (Halle)} {\bf 90} (1909), no.~123.

\bibitem{Goncharov:1998kja}
A.~B. Goncharov, {\it {Multiple polylogarithms, cyclotomy and modular
  complexes}},  {\em Math.Res.Lett.} {\bf 5} (1998) 497--516,
  [\href{https://arxiv.org/abs/1105.2076}{{\tt 1105.2076}}].

\bibitem{Broadhurst:1987ei}
D.~J. Broadhurst, {\it {The Master Two Loop Diagram With Masses}},  {\em Z.
  Phys.} {\bf C47} (1990) 115--124.

\bibitem{Bauberger:1994by}
S.~Bauberger, F.~A. Berends, M.~Bohm, and M.~Buza, {\it {Analytical and
  numerical methods for massive two loop selfenergy diagrams}},  {\em Nucl.
  Phys.} {\bf B434} (1995) 383--407,
  [\href{https://arxiv.org/abs/hep-ph/9409388}{{\tt hep-ph/9409388}}].

\bibitem{Bauberger:1994hx}
S.~Bauberger and M.~Bohm, {\it {Simple one-dimensional integral representations
  for two loop selfenergies: The Master diagram}},  {\em Nucl. Phys.} {\bf
  B445} (1995) 25--48, [\href{https://arxiv.org/abs/hep-ph/9501201}{{\tt
  hep-ph/9501201}}].

\bibitem{Laporta:2004rb}
S.~Laporta and E.~Remiddi, {\it {Analytic treatment of the two loop equal mass
  sunrise graph}},  {\em Nucl. Phys.} {\bf B704} (2005) 349--386,
  [\href{https://arxiv.org/abs/hep-ph/0406160}{{\tt hep-ph/0406160}}].

\bibitem{Kniehl:2005bc}
B.~A. Kniehl, A.~V. Kotikov, A.~Onishchenko, and O.~Veretin, {\it {Two-loop
  sunset diagrams with three massive lines}},  {\em Nucl. Phys.} {\bf B738}
  (2006) 306--316, [\href{https://arxiv.org/abs/hep-ph/0510235}{{\tt
  hep-ph/0510235}}].

\bibitem{Aglietti:2007as}
U.~Aglietti, R.~Bonciani, L.~Grassi, and E.~Remiddi, {\it {The Two loop crossed
  ladder vertex diagram with two massive exchanges}},  {\em Nucl. Phys.} {\bf
  B789} (2008) 45--83, [\href{https://arxiv.org/abs/0705.2616}{{\tt
  0705.2616}}].

\bibitem{Czakon:2008ii}
M.~Czakon and A.~Mitov, {\it {Inclusive Heavy Flavor Hadroproduction in NLO
  QCD: The Exact Analytic Result}},  {\em Nucl. Phys.} {\bf B824} (2010)
  111--135, [\href{https://arxiv.org/abs/0811.4119}{{\tt 0811.4119}}].

\bibitem{Brown:2010bw}
F.~Brown and O.~Schnetz, {\it {A K3 in $\phi^4$}},  {\em Duke Math. J.} {\bf
  161} (2012), no.~10 1817--1862, [\href{https://arxiv.org/abs/1006.4064}{{\tt
  1006.4064}}].

\bibitem{MullerStach:2011ru}
S.~M{\"u}ller-Stach, S.~Weinzierl, and R.~Zayadeh, {\it {A Second-Order
  Differential Equation for the Two-Loop Sunrise Graph with Arbitrary Masses}},
   {\em Commun.Num.Theor.Phys.} {\bf 6} (2012) 203--222,
  [\href{https://arxiv.org/abs/1112.4360}{{\tt 1112.4360}}].

\bibitem{CaronHuot:2012ab}
S.~Caron-Huot and K.~J. Larsen, {\it {Uniqueness of two-loop master contours}},
   {\em JHEP} {\bf 10} (2012) 026, [\href{https://arxiv.org/abs/1205.0801}{{\tt
  1205.0801}}].

\bibitem{Huang:2013kh}
R.~Huang and Y.~Zhang, {\it {On Genera of Curves from High-loop Generalized
  Unitarity Cuts}},  {\em JHEP} {\bf 04} (2013) 080,
  [\href{https://arxiv.org/abs/1302.1023}{{\tt 1302.1023}}].

\bibitem{Brown:2013hda}
F.~Brown and O.~Schnetz, {\it {Modular forms in Quantum Field Theory}},  {\em
  Commun. Num. Theor Phys.} {\bf 07} (2013) 293--325,
  [\href{https://arxiv.org/abs/1304.5342}{{\tt 1304.5342}}].

\bibitem{Nandan:2013ip}
D.~Nandan, M.~F. Paulos, M.~Spradlin, and A.~Volovich, {\it {Star Integrals,
  Convolutions and Simplices}},  {\em JHEP} {\bf 05} (2013) 105,
  [\href{https://arxiv.org/abs/1301.2500}{{\tt 1301.2500}}].

\bibitem{Bloch:2013tra}
S.~Bloch and P.~Vanhove, {\it {The elliptic dilogarithm for the sunset graph}},
   {\em J. Number Theor.} {\bf 148} (2015) 328--364,
  [\href{https://arxiv.org/abs/1309.5865}{{\tt 1309.5865}}].

\bibitem{Adams:2013nia}
L.~Adams, C.~Bogner, and S.~Weinzierl, {\it {The two-loop sunrise graph with
  arbitrary masses}},  {\em J.Math.Phys.} {\bf 54} (2013) 052303,
  [\href{https://arxiv.org/abs/1302.7004}{{\tt 1302.7004}}].

\bibitem{Adams:2014vja}
L.~Adams, C.~Bogner, and S.~Weinzierl, {\it {The two-loop sunrise graph in two
  space-time dimensions with arbitrary masses in terms of elliptic
  dilogarithms}},  {\em J. Math. Phys.} {\bf 55} (2014), no.~10 102301,
  [\href{https://arxiv.org/abs/1405.5640}{{\tt 1405.5640}}].

\bibitem{Adams:2015gva}
L.~Adams, C.~Bogner, and S.~Weinzierl, {\it {The two-loop sunrise integral
  around four space-time dimensions and generalisations of the Clausen and
  Glaisher functions towards the elliptic case}},  {\em J. Math. Phys.} {\bf
  56} (2015), no.~7 072303, [\href{https://arxiv.org/abs/1504.03255}{{\tt
  1504.03255}}].

\bibitem{Adams:2015ydq}
L.~Adams, C.~Bogner, and S.~Weinzierl, {\it {The iterated structure of the
  all-order result for the two-loop sunrise integral}},  {\em J. Math. Phys.}
  {\bf 57} (2016), no.~3 032304, [\href{https://arxiv.org/abs/1512.05630}{{\tt
  1512.05630}}].

\bibitem{Remiddi:2016gno}
E.~Remiddi and L.~Tancredi, {\it {Differential equations and dispersion
  relations for Feynman amplitudes. The two-loop massive sunrise and the kite
  integral}},  {\em Nucl. Phys.} {\bf B907} (2016) 400--444,
  [\href{https://arxiv.org/abs/1602.01481}{{\tt 1602.01481}}].

\bibitem{Primo:2016ebd}
A.~Primo and L.~Tancredi, {\it {On the maximal cut of Feynman integrals and the
  solution of their differential equations}},  {\em Nucl. Phys.} {\bf B916}
  (2017) 94--116, [\href{https://arxiv.org/abs/1610.08397}{{\tt 1610.08397}}].

\bibitem{Bonciani:2016qxi}
R.~Bonciani, V.~Del~Duca, H.~Frellesvig, J.~M. Henn, F.~Moriello, and V.~A.
  Smirnov, {\it {Two-loop planar master integrals for Higgs$\to 3$ partons with
  full heavy-quark mass dependence}},  {\em JHEP} {\bf 12} (2016) 096,
  [\href{https://arxiv.org/abs/1609.06685}{{\tt 1609.06685}}].

\bibitem{Adams:2016xah}
L.~Adams, C.~Bogner, A.~Schweitzer, and S.~Weinzierl, {\it {The kite integral
  to all orders in terms of elliptic polylogarithms}},  {\em {J. Math. Phys.}}
  {\bf 57} (2016), no.~12 122302, [\href{https://arxiv.org/abs/1607.01571}{{\tt
  1607.01571}}].

\bibitem{Passarino:2016zcd}
G.~Passarino, {\it {Elliptic Polylogarithms and Basic Hypergeometric
  Functions}},  {\em Eur. Phys. J.} {\bf C77} (2017), no.~2 77,
  [\href{https://arxiv.org/abs/1610.06207}{{\tt 1610.06207}}].

\bibitem{vonManteuffel:2017hms}
A.~von Manteuffel and L.~Tancredi, {\it {A non-planar two-loop three-point
  function beyond multiple polylogarithms}},  {\em JHEP} {\bf 06} (2017) 127,
  [\href{https://arxiv.org/abs/1701.05905}{{\tt 1701.05905}}].

\bibitem{Ablinger:2017bjx}
J.~Ablinger, J.~Bl{\"u}mlein, A.~De~Freitas, M.~van Hoeij, E.~Imamoglu, C.~G.
  Raab, C.~S. Radu, and C.~Schneider, {\it {Iterated Elliptic and
  Hypergeometric Integrals for Feynman Diagrams}},  {\em J. Math. Phys.} {\bf
  59} (2018), no.~6 062305, [\href{https://arxiv.org/abs/1706.01299}{{\tt
  1706.01299}}].

\bibitem{Chen:2017pyi}
L.-B. Chen, Y.~Liang, and C.-F. Qiao, {\it {NNLO QCD corrections to $\gamma +
  \eta_c(\eta_b)$ exclusive production in electron-positron collision}},  {\em
  JHEP} {\bf 01} (2018) 091, [\href{https://arxiv.org/abs/1710.07865}{{\tt
  1710.07865}}].

\bibitem{Hidding:2017jkk}
M.~Hidding and F.~Moriello, {\it {All orders structure and efficient
  computation of linearly reducible elliptic Feynman integrals}},  {\em JHEP}
  {\bf 01} (2019) 169, [\href{https://arxiv.org/abs/1712.04441}{{\tt
  1712.04441}}].

\bibitem{Bogner:2017vim}
C.~Bogner, A.~Schweitzer, and S.~Weinzierl, {\it {Analytic continuation and
  numerical evaluation of the kite integral and the equal mass sunrise
  integral}},  {\em Nucl. Phys.} {\bf B922} (2017) 528--550,
  [\href{https://arxiv.org/abs/1705.08952}{{\tt 1705.08952}}].

\bibitem{Bourjaily:2017bsb}
J.~L. Bourjaily, A.~J. McLeod, M.~Spradlin, M.~von Hippel, and M.~Wilhelm, {\it
  {Elliptic Double-Box Integrals: Massless Scattering Amplitudes beyond
  Polylogarithms}},  {\em Phys. Rev. Lett.} {\bf 120} (2018), no.~12 121603,
  [\href{https://arxiv.org/abs/1712.02785}{{\tt 1712.02785}}].

\bibitem{Broedel:2017siw}
J.~Broedel, C.~Duhr, F.~Dulat, and L.~Tancredi, {\it {Elliptic polylogarithms
  and iterated integrals on elliptic curves II: an application to the sunrise
  integral}},  {\em Phys. Rev.} {\bf D97} (2018), no.~11 116009,
  [\href{https://arxiv.org/abs/1712.07095}{{\tt 1712.07095}}].

\bibitem{Laporta:2017okg}
S.~Laporta, {\it {High-precision calculation of the 4-loop contribution to the
  electron g-2 in QED}},  {\em Phys. Lett.} {\bf B772} (2017) 232--238,
  [\href{https://arxiv.org/abs/1704.06996}{{\tt 1704.06996}}].

\bibitem{Broedel:2018iwv}
J.~Broedel, C.~Duhr, F.~Dulat, B.~Penante, and L.~Tancredi, {\it {Elliptic
  symbol calculus: from elliptic polylogarithms to iterated integrals of
  Eisenstein series}},  {\em JHEP} {\bf 08} (2018) 014,
  [\href{https://arxiv.org/abs/1803.10256}{{\tt 1803.10256}}].

\bibitem{Broedel:2018qkq}
J.~Broedel, C.~Duhr, F.~Dulat, B.~Penante, and L.~Tancredi, {\it {Elliptic
  Feynman integrals and pure functions}},  {\em JHEP} {\bf 01} (2019) 023,
  [\href{https://arxiv.org/abs/1809.10698}{{\tt 1809.10698}}].

\bibitem{Adams:2018bsn}
L.~Adams, E.~Chaubey, and S.~Weinzierl, {\it {Planar Double Box Integral for
  Top Pair Production with a Closed Top Loop to all orders in the Dimensional
  Regularization Parameter}},  {\em Phys. Rev. Lett.} {\bf 121} (2018), no.~14
  142001, [\href{https://arxiv.org/abs/1804.11144}{{\tt 1804.11144}}].

\bibitem{Adams:2018kez}
L.~Adams, E.~Chaubey, and S.~Weinzierl, {\it {Analytic results for the planar
  double box integral relevant to top-pair production with a closed top loop}},
   {\em JHEP} {\bf 10} (2018) 206,
  [\href{https://arxiv.org/abs/1806.04981}{{\tt 1806.04981}}].

\bibitem{Broedel:2019hyg}
J.~Broedel, C.~Duhr, F.~Dulat, B.~Penante, and L.~Tancredi, {\it {Elliptic
  polylogarithms and Feynman parameter integrals}},  {\em JHEP} {\bf 05} (2019)
  120, [\href{https://arxiv.org/abs/1902.09971}{{\tt 1902.09971}}].

\bibitem{Bogner:2019lfa}
C.~Bogner, S.~M{\"u}ller-Stach, and S.~Weinzierl, {\it {The unequal mass
  sunrise integral expressed through iterated integrals on $\overline{\mathcal
  M}_{1,3}$}},  \href{https://arxiv.org/abs/1907.01251}{{\tt 1907.01251}}.

\bibitem{BrownLevin}
F.~Brown and A.~Levin, {\it {Multiple Elliptic Polylogarithms}},
  \href{https://arxiv.org/abs/1110.6917}{{\tt 1110.6917}}.

\bibitem{Remiddi:2017har}
E.~Remiddi and L.~Tancredi, {\it {An Elliptic Generalization of Multiple
  Polylogarithms}},  {\em Nucl. Phys.} {\bf B925} (2017) 212--251,
  [\href{https://arxiv.org/abs/1709.03622}{{\tt 1709.03622}}].

\bibitem{Broedel:2017kkb}
J.~Broedel, C.~Duhr, F.~Dulat, and L.~Tancredi, {\it {Elliptic polylogarithms
  and iterated integrals on elliptic curves. Part I: general formalism}},  {\em
  JHEP} {\bf 05} (2018) 093, [\href{https://arxiv.org/abs/1712.07089}{{\tt
  1712.07089}}].

\bibitem{ManinModular}
Y.~I. Manin, {\it {Iterated integrals of modular forms and noncommutative
  modular symbols}},  in {\em Algebraic geometry and number theory}, vol.~253
  of {\em Progr. Math.}, (Boston), pp.~565--597, Birkh\"auser Boston, 2006.
\newblock \href{https://arxiv.org/abs/math/0502576}{{\tt math/0502576}}.

\bibitem{Brown:mmv}
F.~Brown, {\it Multiple modular values and the relative completion of the
  fundamental group of $\mathcal{M}_{1,1}$},
  \href{https://arxiv.org/abs/1407.5167v4}{{\tt 1407.5167v4}}.

\bibitem{Adams:2017ejb}
L.~Adams and S.~Weinzierl, {\it {Feynman integrals and iterated integrals of
  modular forms}},  {\em Commun. Num. Theor. Phys.} {\bf 12} (2018) 193--251,
  [\href{https://arxiv.org/abs/1704.08895}{{\tt 1704.08895}}].

\bibitem{Broedel:2014vla}
J.~Broedel, C.~R. Mafra, N.~Matthes, and O.~Schlotterer, {\it {Elliptic
  multiple zeta values and one-loop superstring amplitudes}},  {\em JHEP} {\bf
  07} (2015) 112, [\href{https://arxiv.org/abs/1412.5535}{{\tt 1412.5535}}].

\bibitem{Broedel:2015hia}
J.~Broedel, N.~Matthes, and O.~Schlotterer, {\it {Relations between elliptic
  multiple zeta values and a special derivation algebra}},  {\em J. Phys.} {\bf
  A49} (2016), no.~15 155203, [\href{https://arxiv.org/abs/1507.02254}{{\tt
  1507.02254}}].

\bibitem{Broedel:2017jdo}
J.~Broedel, N.~Matthes, G.~Richter, and O.~Schlotterer, {\it {Twisted elliptic
  multiple zeta values and non-planar one-loop open-string amplitudes}},  {\em
  J. Phys.} {\bf A51} (2018), no.~28 285401,
  [\href{https://arxiv.org/abs/1704.03449}{{\tt 1704.03449}}].

\bibitem{Broedel:2018izr}
J.~Broedel, O.~Schlotterer, and F.~Zerbini, {\it {From elliptic multiple zeta
  values to modular graph functions: open and closed strings at one loop}},
  {\em JHEP} {\bf 01} (2019) 155, [\href{https://arxiv.org/abs/1803.00527}{{\tt
  1803.00527}}].

\bibitem{Bloch:2014qca}
S.~Bloch, M.~Kerr, and P.~Vanhove, {\it {A Feynman integral via higher normal
  functions}},  {\em Compos. Math.} {\bf 151} (2015) 2329--2375,
  [\href{https://arxiv.org/abs/1406.2664}{{\tt 1406.2664}}].

\bibitem{Bloch:2016izu}
S.~Bloch, M.~Kerr, and P.~Vanhove, {\it {Local mirror symmetry and the sunset
  Feynman integral}},  {\em Adv. Theor. Math. Phys.} {\bf 21} (2016), no.~6
  [\href{https://arxiv.org/abs/1601.08181}{{\tt 1601.08181}}].

\bibitem{Broadhurst:2016myo}
D.~Broadhurst, {\it {Feynman integrals, L-series and Kloosterman moments}},
  {\em Commun. Num. Theor. Phys.} {\bf 10} (2016) 527--569,
  [\href{https://arxiv.org/abs/1604.03057}{{\tt 1604.03057}}].

\bibitem{Bourjaily:2018yfy}
J.~L. Bourjaily, A.~J. McLeod, M.~von Hippel, and M.~Wilhelm, {\it {A (Bounded)
  Bestiary of Feynman Integral Calabi-Yau Geometries}},  {\em Phys. Rev. Lett.}
  {\bf 122} (2019), no.~3 031601, [\href{https://arxiv.org/abs/1810.07689}{{\tt
  1810.07689}}].

\bibitem{Bourjaily:2018ycu}
J.~L. Bourjaily, Y.-H. He, A.~J. Mcleod, M.~Von~Hippel, and M.~Wilhelm, {\it
  {Traintracks through Calabi-Yau Manifolds: Scattering Amplitudes beyond
  Elliptic Polylogarithms}},  {\em Phys. Rev. Lett.} {\bf 121} (2018), no.~7
  071603, [\href{https://arxiv.org/abs/1805.09326}{{\tt 1805.09326}}].

\bibitem{verrill1996}
H.~A. Verrill, {\it Root lattices and pencils of varieties},  {\em J. Math.
  Kyoto Univ.} {\bf 36} (1996), no.~2 423--446.

\bibitem{Primo:2017ipr}
A.~Primo and L.~Tancredi, {\it {Maximal cuts and differential equations for
  Feynman integrals. An application to the three-loop massive banana graph}},
  {\em Nucl. Phys.} {\bf B921} (2017) 316--356,
  [\href{https://arxiv.org/abs/1704.05465}{{\tt 1704.05465}}].

\bibitem{Mastrolia:2002tv}
P.~Mastrolia and E.~Remiddi, {\it {The Analytic value of a three loop sunrise
  graph in a particular kinematical configuration}},  {\em Nucl. Phys.} {\bf
  B657} (2003) 397--406, [\href{https://arxiv.org/abs/hep-ph/0211451}{{\tt
  hep-ph/0211451}}].

\bibitem{Chetyrkin:1981qh}
K.~Chetyrkin and F.~Tkachov, {\it {Integration by Parts: The Algorithm to
  Calculate beta Functions in 4 Loops}},  {\em Nucl.Phys.} {\bf B192} (1981)
  159--204.

\bibitem{Tkachov:1981wb}
F.~V. Tkachov, {\it {A Theorem on Analytical Calculability of Four Loop
  Renormalization Group Functions}},  {\em Phys. Lett.} {\bf B100} (1981)
  65--68.

\bibitem{Tarasov:1996br}
O.~V. Tarasov, {\it {Connection between Feynman integrals having different
  values of the space-time dimension}},  {\em Phys. Rev.} {\bf D54} (1996)
  6479--6490, [\href{https://arxiv.org/abs/hep-th/9606018}{{\tt
  hep-th/9606018}}].

\bibitem{Smirnov:1999wz}
V.~A. Smirnov and O.~L. Veretin, {\it {Analytical results for dimensionally
  regularized massless on-shell double boxes with arbitrary indices and
  numerators}},  {\em Nucl. Phys.} {\bf B566} (2000) 469--485,
  [\href{https://arxiv.org/abs/hep-ph/9907385}{{\tt hep-ph/9907385}}].

\bibitem{Anastasiou:1999bn}
C.~Anastasiou, E.~W.~N. Glover, and C.~Oleari, {\it {The two-loop scalar and
  tensor pentabox graph with light-like legs}},  {\em Nucl. Phys.} {\bf B575}
  (2000) 416--436, [\href{https://arxiv.org/abs/hep-ph/9912251}{{\tt
  hep-ph/9912251}}]. [Erratum: Nucl. Phys.B585,763(2000)].

\bibitem{Anastasiou:2000mf}
C.~Anastasiou, T.~Gehrmann, C.~Oleari, E.~Remiddi, and J.~B. Tausk, {\it {The
  Tensor reduction and master integrals of the two loop massless crossed box
  with lightlike legs}},  {\em Nucl. Phys.} {\bf B580} (2000) 577--601,
  [\href{https://arxiv.org/abs/hep-ph/0003261}{{\tt hep-ph/0003261}}].

\bibitem{Anastasiou:2000kp}
C.~Anastasiou, J.~B. Tausk, and M.~E. Tejeda-Yeomans, {\it {The On-shell
  massless planar double box diagram with an irreducible numerator}},  {\em
  Nucl. Phys. Proc. Suppl.} {\bf 89} (2000) 262--267,
  [\href{https://arxiv.org/abs/hep-ph/0005328}{{\tt hep-ph/0005328}}].

\bibitem{Lee:2009dh}
R.~N. Lee, {\it {Space-time dimensionality D as complex variable: Calculating
  loop integrals using dimensional recurrence relation and analytical
  properties with respect to D}},  {\em Nucl. Phys.} {\bf B830} (2010)
  474--492, [\href{https://arxiv.org/abs/0911.0252}{{\tt 0911.0252}}].

\bibitem{Kotikov:1990kg}
A.~V. Kotikov, {\it {Differential equations method: New technique for massive
  Feynman diagrams calculation}},  {\em Phys. Lett.} {\bf B254} (1991)
  158--164.

\bibitem{Kotikov:1991hm}
A.~V. Kotikov, {\it {Differential equations method: The Calculation of vertex
  type Feynman diagrams}},  {\em Phys. Lett.} {\bf B259} (1991) 314--322.

\bibitem{Kotikov:1991pm}
A.~V. Kotikov, {\it {Differential equation method: The Calculation of N point
  Feynman diagrams}},  {\em Phys. Lett.} {\bf B267} (1991) 123--127.

\bibitem{Gehrmann:1999as}
T.~Gehrmann and E.~Remiddi, {\it {Differential equations for two loop four
  point functions}},  {\em Nucl.Phys.} {\bf B580} (2000) 485--518,
  [\href{https://arxiv.org/abs/hep-ph/9912329}{{\tt hep-ph/9912329}}].

\bibitem{Gehrmann:2000zt}
T.~Gehrmann and E.~Remiddi, {\it {Two loop master integrals for $\gamma^*\to 3$
  jets: The Planar topologies}},  {\em Nucl.Phys.} {\bf B601} (2001) 248--286,
  [\href{https://arxiv.org/abs/hep-ph/0008287}{{\tt hep-ph/0008287}}].

\bibitem{Frellesvig:2017aai}
H.~Frellesvig and C.~G. Papadopoulos, {\it {Cuts of Feynman Integrals in Baikov
  representation}},  {\em JHEP} {\bf 04} (2017) 083,
  [\href{https://arxiv.org/abs/1701.07356}{{\tt 1701.07356}}].

\bibitem{Bosma:2017ens}
J.~Bosma, M.~Sogaard, and Y.~Zhang, {\it {Maximal Cuts in Arbitrary
  Dimension}},  {\em JHEP} {\bf 08} (2017) 051,
  [\href{https://arxiv.org/abs/1704.04255}{{\tt 1704.04255}}].

\bibitem{joyce}
G.~Joyce, {\it {On the simple cubic lattice Green function}},  {\em
  Transactions of the Royal Society of London, Mathematical and Physical
  Sciences} {\bf 1973} (273) 583--610.

\bibitem{Sabry}
A.~Sabry, {\it {Fourth order spectral functions for the electron propagator}},
  {\em Nucl. Phys.} {\bf 33} (1962), no.~17 401--430.

\bibitem{Adams:2013kgc}
L.~Adams, C.~Bogner, and S.~Weinzierl, {\it {The two-loop sunrise graph with
  arbitrary masses}},  {\em J. Math. Phys.} {\bf 54} (2013) 052303,
  [\href{https://arxiv.org/abs/1302.7004}{{\tt 1302.7004}}].

\bibitem{Maier}
R.~S. {Maier}, {\it {On Rationally Parametrized Modular Equations}},  {\em
  ArXiv Mathematics e-prints} (Nov., 2006)
  [\href{https://arxiv.org/abs/math/0611041}{{\tt math/0611041}}].

\bibitem{Broedel:2018rwm}
J.~Broedel, C.~Duhr, F.~Dulat, B.~Penante, and L.~Tancredi, {\it {From modular
  forms to differential equations for Feynman integrals}},  in {\em {KMPB
  Conference: Elliptic Integrals, Elliptic Functions and Modular Forms in
  Quantum Field Theory Zeuthen, Germany, October 23-26, 2017}}, 2018.
\newblock \href{https://arxiv.org/abs/1807.00842}{{\tt 1807.00842}}.

\bibitem{Pollack}
A.~Pollack, {\it {Relations between derivations arising from modular forms}}, .

\bibitem{Davies:2019nhm}
J.~Davies, R.~Gr{\"o}ber, A.~Maier, T.~Rauh, and M.~Steinhauser, {\it {Top
  quark mass dependence of the Higgs-gluon form factor at three loops}},
  \href{https://arxiv.org/abs/1906.00982}{{\tt 1906.00982}}.

\bibitem{regions1}
M.~Beneke and V.~A. Smirnov, {\it {Asymptotic expansion of Feynman integrals
  near threshold}},  {\em Nucl. Phys. B} {\bf 522} (1998) 321.

\bibitem{regions2}
V.~A. Smirnov, {\em {Applied asymptotic expansions in momenta and masses}},
  vol.~177.
\newblock Springer Tracts Mod. Phys., 2002.

\bibitem{Czakon:2005rk}
M.~Czakon, {\it {Automatized analytic continuation of Mellin-Barnes
  integrals}},  {\em Comput. Phys. Commun.} {\bf 175} (2006) 559--571,
  [\href{https://arxiv.org/abs/hep-ph/0511200}{{\tt hep-ph/0511200}}].

\bibitem{Smirnov:2009up}
A.~V. Smirnov and V.~A. Smirnov, {\it {On the Resolution of Singularities of
  Multiple Mellin-Barnes Integrals}},  {\em Eur. Phys. J.} {\bf C62} (2009)
  445--449, [\href{https://arxiv.org/abs/0901.0386}{{\tt 0901.0386}}].

\bibitem{mbasymptotics}
M.~Czakon, {\it \verb+https://mbtools.hepforge.org/+}, .

\end{thebibliography}\endgroup

\end{document}